\input lanlmac
\def\href#1#2{{#2}}
\def\hhref#1{{#1}}
\input epsf.tex

\overfullrule=0mm

\newcount\figno
\figno=0
\def\fig#1#2#3{
\par\begingroup\parindent=0pt\leftskip=1cm\rightskip=1cm\parindent=0pt
\baselineskip=11pt
\global\advance\figno by 1
\midinsert
\epsfxsize=#3
\centerline{\epsfbox{#2}}
\vskip 12pt
{\bf Fig.\ \the\figno:} #1\par
\endinsert\endgroup\par
}
\def\figlabel#1{\xdef#1{\the\figno}}
\def\encadremath#1{\vbox{\hrule\hbox{\vrule\kern8pt\vbox{\kern8pt
\hbox{$\displaystyle #1$}\kern8pt}
\kern8pt\vrule}\hrule}}


\def\IR{\relax{\rm I\kern-.18em R}}
\font\cmss=cmss10 \font\cmsss=cmss10 at 7pt

\def\q#1{\left[#1\right]_x}

\font\cmss=cmss10 \font\cmsss=cmss10 at 7pt
\def\IZ{\relax\ifmmode\mathchoice
{\hbox{\cmss Z\kern-.4em Z}}{\hbox{\cmss Z\kern-.4em Z}}
{\lower.9pt\hbox{\cmsss Z\kern-.4em Z}}
{\lower1.2pt\hbox{\cmsss Z\kern-.4em Z}}\else{\cmss Z\kern-.4em Z}\fi}
\def\IN{\relax{\rm I\kern-.18em N}}
\def\b{\circ}
\def\n{\bullet}

\def\gbbbb{\Gamma_4^{\hbox{$\scriptstyle \b \b$}\kern -8.2pt
\raise -4pt \hbox{$\scriptstyle \b \b$}}}
\def\gnnnn{\Gamma_4^{\hbox{$\scriptstyle \n \n$}\kern -8.2pt  
\raise -4pt \hbox{$\scriptstyle \n \n$}}}
\def\gnnnnnn{\Gamma_6^{\hbox{$\scriptstyle \n \n \n$}\kern -12.3pt
\raise -4pt \hbox{$\scriptstyle \n \n \n$}}}
\def\gbbbbbb{\Gamma_6^{\hbox{$\scriptstyle \b \b \b$}\kern -12.3pt
\raise -4pt \hbox{$\scriptstyle \b \b \b$}}}
\def\gbbbbc{\Gamma_{4\, c}^{\hbox{$\scriptstyle \b \b$}\kern -8.2pt
\raise -4pt \hbox{$\scriptstyle \b \b$}}}
\def\gnnnnc{\Gamma_{4\, c}^{\hbox{$\scriptstyle \n \n$}\kern -8.2pt
\raise -4pt \hbox{$\scriptstyle \n \n$}}}
\def\Rbud#1{{\cal R}_{#1}^{-\kern-1.5pt\blacktriangleright}}
\def\Rleaf#1{{\cal R}_{#1}^{-\kern-1.5pt\vartriangleright}}
\def\Rbudb#1{{\cal R}_{#1}^{\circ\kern-1.5pt-\kern-1.5pt\blacktriangleright}}
\def\Rleafb#1{{\cal R}_{#1}^{\circ\kern-1.5pt-\kern-1.5pt\vartriangleright}}
\def\Rbudn#1{{\cal R}_{#1}^{\bullet\kern-1.5pt-\kern-1.5pt\blacktriangleright}}
\def\Rleafn#1{{\cal R}_{#1}^{\bullet\kern-1.5pt-\kern-1.5pt\vartriangleright}}
\def\Wleaf#1{{\cal W}_{#1}^{-\kern-1.5pt\vartriangleright}}
\def\Cleaf{{\cal C}^{-\kern-1.5pt\vartriangleright}}
\def\Cbud{{\cal C}^{-\kern-1.5pt\blacktriangleright}}
\def\Crleaf{{\cal C}^{-\kern-1.5pt\circledR}}


\magnification=\magstep1
\baselineskip=12pt
\hsize=6.3truein
\vsize=8.7truein
 at 8truept
 at 8truept
 at 10truept

\font\bigrm=cmr12 at 14pt
\centerline{\bigrm The three-point function of planar quadrangulations}

\bigskip\bigskip

\centerline{J. Bouttier and E. Guitter}
  \smallskip
  \centerline{Institut de Physique Th\'eorique}
  \centerline{CEA, IPhT, F-91191 Gif-sur-Yvette, France}
  \centerline{CNRS, URA 2306}
\centerline{\tt jeremie.bouttier@cea.fr}
\centerline{\tt emmanuel.guitter@cea.fr}

  \bigskip

     \bigskip\bigskip

     \centerline{\bf Abstract}
     \smallskip
     {\narrower\noindent
We compute the generating function of random planar quadrangulations
with three marked vertices at prescribed pairwise distances. In the 
scaling limit of large quadrangulations, this discrete three-point
function converges to a simple universal scaling function, which is
the continuous three-point function of pure 2D quantum gravity. We give
explicit expressions for this universal three-point function both in the
grand-canonical and canonical ensembles. Various limiting regimes
are studied when some of the distances become large or small. 
By considering the case where the marked vertices are aligned, we 
also obtain the probability law for the number of 
geodesic points, namely vertices that lie on a geodesic path between
two given vertices, and at prescribed distances from these vertices.
\par}

     \bigskip

\nref\QGRA{V. Kazakov, {\it Bilocal regularization of models of random
surfaces}, Phys. Lett. {\bf B150} (1985) 282-284; F. David, {\it Planar
diagrams, two-dimensional lattice gravity and surface models},
Nucl. Phys. {\bf B257} (1985) 45-58; J. Ambj\o rn, B. Durhuus and J. Fr\"ohlich,
{\it Diseases of triangulated random surface models and possible cures},
Nucl. Phys. {\bf B257}(1985) 433-449; V. Kazakov, I. Kostov and A. Migdal
{\it Critical properties of randomly triangulated planar random surfaces},
Phys. Lett. {\bf B157} (1985) 295-300.}
\nref\TUT{W. Tutte,
{\it A Census of planar triangulations} Canad. J. of Math. {\bf 14} 
(1962) 21-38;
{\it A Census of Hamiltonian polygons} Canad. J. of Math. 
{\bf 14} (1962) 402-417;
{\it A Census of slicings}, Canad. J. of Math. {\bf 14} (1962) 708-722;
{\it A Census of Planar Maps}, Canad. J. of Math. {\bf 15} (1963) 249-271.
}
\nref\BIPZ{E. Br\'ezin, C. Itzykson, G. Parisi and J.-B. Zuber, {\it Planar
Diagrams}, Comm. Math. Phys. {\bf 59} (1978) 35-51.}
\nref\DGZ{P. Di Francesco, P. Ginsparg and J. Zinn--Justin, {\it 2D Gravity and Random Matrices},
Physics Reports {\bf 254} (1995) 1-131.}
\nref\SCH{G. Schaeffer, {\it Bijective census and random
generation of Eulerian planar maps}, Electronic
Journal of Combinatorics, vol. {\bf 4} (1997) R20; see also
{\it Conjugaison d'arbres
et cartes combinatoires al\'eatoires}, PhD Thesis, Universit\'e 
Bordeaux I (1998).}
\nref\CONST{M. Bousquet-M\'elou and G. Schaeffer,
{\it Enumeration of planar constellations}, Adv. in Applied Math.,
{\bf 24} (2000) 337-368.}
\nref\CENSUS{J. Bouttier, P. Di Francesco and E. Guitter, {\it Census of planar
maps: from the one-matrix model solution to a combinatorial proof},
Nucl. Phys. {\bf B645}[PM] (2002) 477-499, arXiv:cond-mat/0207682.}
\nref\BMS{M. Bousquet-M\'elou and G. Schaeffer,{\it The degree distribution
in bipartite planar maps: application to the Ising model},
arXiv:math.CO/0211070.}
\nref\HPCOM{J. Bouttier, P. Di Francesco and E. Guitter. 
{\it Combinatorics of Hard Particles
on Planar Graphs}, Nucl.Phys. {\bf B655} (2003) 313-341, 
arXiv:cond-mat/0211168.}
\nref\HObipar{J. Bouttier, P. Di Francesco and E. Guitter. 
{\it Combinatorics of bicubic maps 
with hard particles}, J.Phys. A: Math.Gen. {\bf 38} 
(2005) 4529-4560, arXiv:math.CO/0501344.}
\nref\MS{M. Marcus and G. Schaeffer, {\it Une bijection simple pour les
cartes orientables}, (2001), available at 
\hhref{http://www.lix.polytechnique.fr/Labo/Gilles.Schaeffer/Biblio/};
see also G. Chapuy, M. Marcus and G. Schaeffer, 
{\it A bijection for rooted maps on orientable surfaces}, 
arXiv: math-CO/0712.3649 and G. Schaeffer, 
{\it Conjugaison d'arbres
et cartes combinatoires al\'eatoires}, PhD Thesis, Universit\'e
Bordeaux I (1998).}
\nref\CS{P. Chassaing and G. Schaeffer, {\it Random Planar Lattices and 
Integrated SuperBrownian Excursion}, 
Probability Theory and Related Fields {\bf 128(2)} (2004) 161-212, 
arXiv:math.CO/0205226.}
\nref\MOB{J. Bouttier, P. Di Francesco and E. Guitter. {\it 
Planar maps as labeled mobiles},
Elec. Jour. of Combinatorics {\bf 11} (2004) R69, arXiv:math.CO/0405099.}
\nref\FOMAP{J. Bouttier, P. Di Francesco and E. Guitter. {\it Blocked edges 
on Eulerian maps and mobiles: Application to spanning trees, hard particles 
and the Ising model}, 	J. Phys. A: Math. Theor. {\bf 40} (2007) 7411-7440, 
arXiv:math.CO/0702097.}
\nref\AW{J. Ambj\o rn and Y. Watabiki, {\it Scaling in quantum gravity},
Nucl.Phys. {\bf B445} (1995) 129-144.}
\nref\AJW{J. Ambj\o rn, J. Jurkiewicz and Y. Watabiki,
{\it On the fractal structure of two-dimensional quantum gravity},
Nucl.Phys. {\bf B454} (1995) 313-342.}
\nref\GEOD{J. Bouttier, P. Di Francesco and E. Guitter, {\it Geodesic
distance in planar graphs}, Nucl. Phys. {\bf B663}[FS] (2003) 535-567, 
arXiv:cond-mat/0303272.}
\nref\STATGEOD{J. Bouttier and E. Guitter, {\it Statistics of
geodesics in large quadrangulations}, J. Phys. A: Math. Theor. {\bf 41} 
(2008) 145001 (30pp), arXiv:math-ph/0712.2160.}
\nref\Mier{G. Miermont, {\it Tessellations of random maps of arbitrary
genus}, arXiv:math.PR/0712.3688.}
\newsec{Introduction}
\subsec{The problem}
Maps are fundamental objects of discrete mathematics, usually defined as
cellular embeddings of graphs into surfaces. They raise many interesting 
combinatorial and probabilistic questions. The understanding of their 
statistical properties is also relevant to physics, where random maps 
provide discrete models for fluctuating surfaces, for instance in the 
context of two-dimensional quantum gravity \QGRA. Many results have been 
obtained over the years for the {\it enumeration} of various families of maps, 
including maps carrying spins or particles. Besides Tutte's original approach 
using some recursive decomposition of the maps \TUT, several new techniques of
enumeration were introduced over the years, using for instance random 
matrix integrals [\xref\BIPZ,\xref\DGZ] or, more recently, bijections
with decorated trees like ``blossom trees'' [\xref\SCH-\xref\HObipar]
or ``well-labeled trees'' [\xref\MS-\xref\FOMAP]. So far, most enumeration
results dealt however with global properties of the maps, consisting in 
simply counting these maps or computing {\it global} correlations 
functions for some observables, obtained by averaging over the position of 
the points where these observables are measured (see \DGZ). To fully 
understand the structure of random maps, it is however necessary
to have access to {\it local} correlations, in which one controls the distance 
between the points where the measure takes place [\xref\AW,\xref\AJW]. 
Very little is known at this time about local correlations but
many results in this direction should be within reach by a proper use of the 
above-mentioned bijection with well-labeled trees. Indeed, in this approach, 
the coding of maps into trees makes explicit reference to the graph distance 
to some origin vertex, thus keeping track in the enumeration of a number 
of graph distances between vertices. A first progress was achieved 
in Ref.~\GEOD, where the so-called canonical {\it two-point function} of 
quadrangulations was computed. This function gives the average number of
pairs of vertices at prescribed distance from each other in the
ensemble of quadrangulations (i.e.\ maps with only tetravalent faces)
with a fixed number of faces. In particular, 
for large quadrangulations and at large distances, a sensible scaling
limit can be obtained where the two-point function converges 
to some universal scaling function, which is the two-point function 
of pure 2D quantum gravity [\xref\AW,\xref\GEOD]. Another recent
progress was made in Ref.\STATGEOD, where the question of geodesic paths 
between two vertices in random quadrangulations was addressed, and a number 
of results were obtained on the actual dependence of the statistics
of geodesics on the distance between their endpoints.

In this paper, we address the question of the {\it three-point function}
of planar quadrangulations, consisting in enumerating triply-pointed 
quadrangulations,
i.e.\ quadrangulations with three marked vertices {\it at prescribed pairwise 
distances}. Our main result is an explicit formula for the generating 
function of such triply-pointed quadrangulations counted with a weight 
$g$ per face. Our approach relies on an extension by Miermont of the bijection 
with well-labeled trees which allows to treat the case of multiply-pointed 
quadrangulations \Mier\ via a coding by more general well-labeled maps, 
which can then be enumerated. 
Considering the scaling limit of large quadrangulations and large distances, 
we then give explicit expressions for the universal scaling form of this 
three-point function, in both grand-canonical and canonical ensembles.
In the language of 2D quantum gravity, this constitutes the continuous 
three-point function for the universality class of the so-called pure 
gravity, with the planar topology. We discuss here its main properties. 
 
The paper is organized as follows: in section 1.2, we start by briefly 
recalling some known results about the two-point function of quadrangulations,
and its continuum limit. We then present in section 2 our results for
the three-point function. We give in section 2.1 the explicit formula
for the generating function $G(d_{12},d_{23},d_{31};g)$ with a weight
$g$ per face of planar quadrangulations with three distinct vertices, 
distinguished as $1$, $2$ and $3$, at prescribed pairwise distances 
$d_{12}$, $d_{23}$ and $d_{31}$. We then derive in section 2.2 its
universal scaling form both in the grand-canonical ensemble with a fixed
``cosmological constant'' and in the canonical ensemble of
maps with a fixed, but large number $n$ of faces. The grand-canonical 
three-point function is obtained by letting $g$ tend to its 
critical value $1/12$ and considering distances scaling
as $(1/12-g)^{-1/4}$. As for the canonical three-point function,
it is obtained by letting the size $n$ of the quadrangulations tend
to infinity, with distances scaling as $n^{1/4}$. 
Various limits of the canonical three-point function are discussed 
in section 2.3, corresponding to the cases when one or several of the
(rescaled) distances become large or small. Sections 3 and 4 are
devoted to the actual derivation of our main formula for 
$G(d_{12},d_{23},d_{31};g)$. We first recall in section 3.1 the Miermont 
bijection between, on the one hand, multiply-pointed
quadrangulations with sources and delays and, on the other hand, 
well-labeled maps.
We show in section 3.2 how to use this bijection in the case
of triply-pointed maps and how to keep track of all pairwise distances
by a proper choice of delays. This allows us to reduce our enumeration
problem to that of counting particular well-labeled maps with three faces, 
with a number of constraints on their labels. 
The generating function for such maps is finally obtained
in section 4, where we make use of three main building blocks: the already
known generating function for well-labeled trees, recalled in section 4.1, 
a new generating function for properly weighted Motzkin paths 
describing chains of such trees corresponding to well-labeled maps 
with two faces, explicited in section 4.2 and another new generating
function for so-called Y-diagrams appearing in a decomposition of
well-labeled maps with three faces, explicited in section 4.3, 
where we finally establish our formula for $G(d_{12},d_{23},d_{31};g)$.
An alternative derivation of the above generating function for weighted
Motzkin path is presented in Appendix A. 
We discuss in section 4.4 a number of applications of our formula 
corresponding to the so-called (non-universal) ``local limit'' of large 
quadrangulations, where we keep the distances finite but let
the number $n$ of faces tend to infinity.
We discuss in details the case where one of the three marked vertices
is a {\it geodesic point}, i.e.\ lies on a geodesic path between the 
two others. We give in particular the probability law for the
number of geodesic points at fixed distances from two marked vertices.
We finally discuss possible extensions of our results and 
conclude in section 5.

\subsec{The two-point function of quadrangulations}

Before we present our results for the three-point function, it is
useful to recall briefly some known facts about the two-point function
of quadrangulations. This function is obtained by considering configurations 
of doubly-pointed planar quadrangulations, i.e.\ planar 
quadrangulations with {\it two 
marked distinct and distinguished vertices}. As customary, every configuration
is counted with a weight $g$ per face and some inverse symmetry factor,
equal to the inverse of the order of its automorphism group. 
The generating function $G(i;g)$ of such weighted doubly-pointed 
quadrangulations where the two marked points are at distance $i$ from each 
other was found to be \MOB:
\eqn\generi{G(i;g)=\left\{\matrix{&
\displaystyle{\log \left({R_i\over R_{i-1}}\right)} &\ {\rm for}\  i\geq 2 
\cr & & \cr
& \displaystyle{\log R_1} & \ {\rm for}\ i=1\cr}\right.}
with $R_i$ given by \GEOD:
\eqn\Rdei{R_i= R {\q{i}\, \q{i+3} \over
\q{i+1}\, \q{i+2}}\ ,\qquad i\geq 1\ .}
Here and throughout the paper, we use the shorthand notation:
\eqn\defq{\q{i}\equiv {1-x^i\over 1-x}\ .}
The quantities $R$ and $x$ in \Rdei\ are power series in $g$ determined by
\eqn\xRtog{g={x+1+{1\over x}\over \left(x+4+{1\over x}\right)^2}\ ,\qquad
R={x+4+{1\over x}\over x+1+{1\over x}}\ ,} 
so that we have
\eqn\Rxexplicit{\eqalign{R &= {1-\sqrt{1-12g}\over 6 g}\ ,\cr
x &={1-24g-\sqrt{1-12g}+\sqrt{6}\sqrt{72g^2+6g\sqrt{1-12g}-1}\over
2(6g+\sqrt{1-12g}-1)}\ .\cr}}
The quantity $G(i;g)$ constitutes the discrete ``grand-canonical'' 
two-point function of quadrangulations and corresponds to an ensemble
of quadrangulations with a varying number of faces, 
governed by $g$. We may instead 
decide to fix the number $n$ of faces of the quadrangulation. The
corresponding partition function can be obtained via a contour integral 
in $g$ around $0$ as:
\eqn\canG{G(i;g)\vert_{g^n}={1\over 2{\rm i}\pi} \oint {dg\over  g^{n+1}}
G(i;g)\ .}
This quantity constitutes the discrete ``canonical'' two-point function 
of quadrangulations.
\fig{The universal canonical two-point function, given by (a) the
probability density function $\rho(D)$, and (b) the corresponding 
distribution function $\Phi(D)$.}{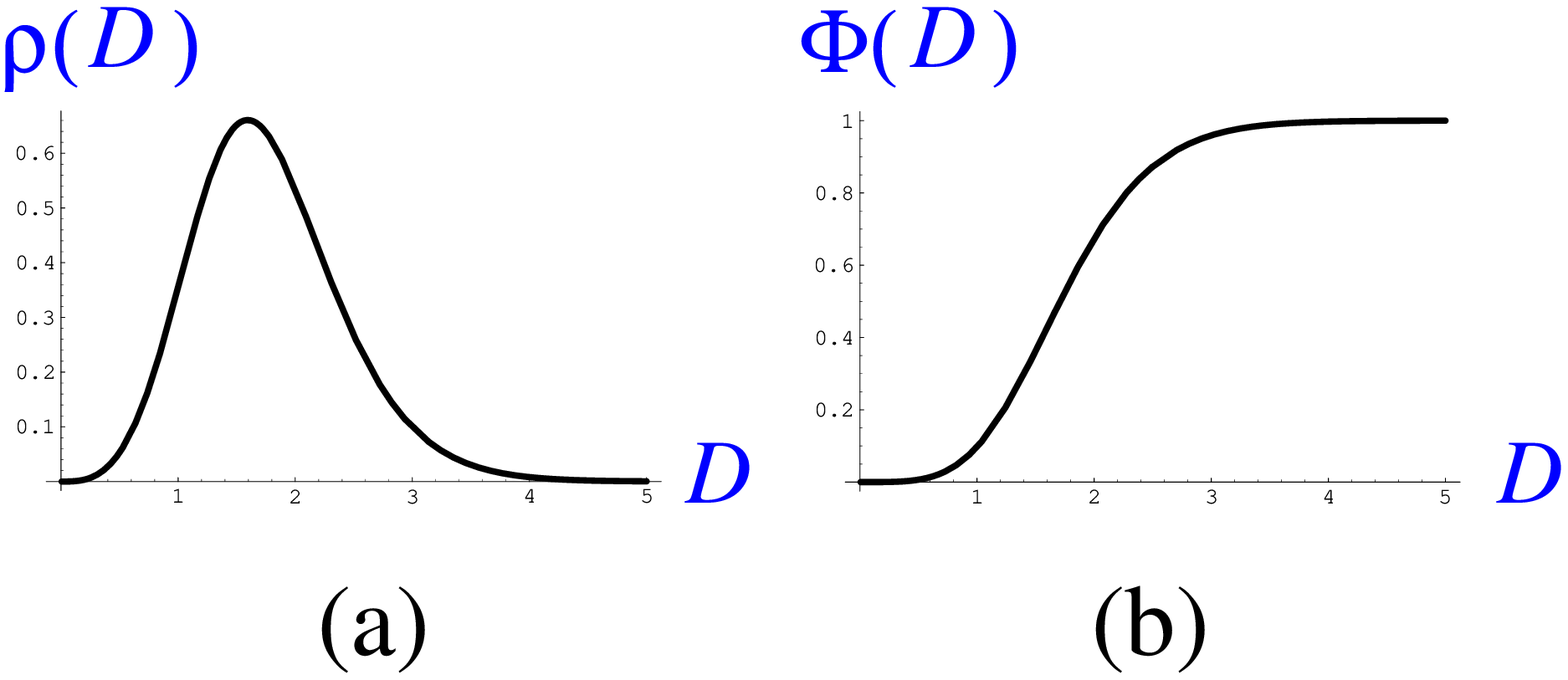}{12.cm}
\figlabel\rhophi
A sensible continuum limit is reached by letting $g$ approach 
its critical value $1/12$ and jointly taking a large distance $i$, 
with the following scaling:
\eqn\scalone{g={1\over 12}\left(1- \Lambda\, \epsilon\right)\ , \qquad
i =D \epsilon^{-1/4}\ ,}
where $\epsilon\to 0$ and where $\Lambda$ is the so-called 
``cosmological constant''. 
In this limit, we have
\eqn\valFtwo{\log\left({R_i\over 2}\right) \sim \epsilon^{1/2} 
{\cal F}\left(D;\sqrt{3/2}\Lambda^{1/4}\right)
\quad {\rm with}\quad {\cal F}(D;\alpha)= 
-{2 \alpha^2 \over 3} \left(1+{3
\over \sinh^2(\alpha D)}\right)}
so that 
\eqn\valGtwo{G(i;g)\sim \epsilon^{3/4} {\cal G}\left(D;
 \sqrt{3/2}\Lambda^{1/4}\right) \quad  
{\rm with}\quad {\cal G}(D;\alpha)= \partial_D {\cal F}(D;\alpha)
= 4 \alpha^3 {\cosh(\alpha D)\over \sinh^3(\alpha D)}\ .}
This constitutes the continuous two-point function in the grand-canonical 
ensemble with cosmological constant $\Lambda$ [\xref\AW,\xref\AJW]. 
Again, we may instead 
consider the canonical ensemble where the number of faces $n$ of the 
quadrangulations is now fixed to a large value. A sensible limit is reached 
by letting $n$ tend to infinity with the distance between the marked vertices
scaling as $n^{1/4}$. The integral \canG\ translates at large $n$ via a 
saddle point estimate into an integral over {\it real values}
of some parameter $\xi$, upon setting
\eqn\scalonebis{g={1\over 12}\left(1+{\xi^2\over n}\right)\ , \qquad
i =D n^{1/4}\ .}
Using Eq.~\valGtwo\ with $\epsilon=1/n$ and $\Lambda=-\xi^2$, 
and performing a proper normalization of $G(i;g)\vert_{g^n}$ by the number 
of doubly-pointed quadrangulations with size $n$, we then obtain 
a normalized probability density 
\eqn\conttwo{\rho(D)= {2\over {\rm i} \sqrt{\pi}} \int_{-\infty}^{\infty} 
d\xi\ \xi\, e^{-\xi^2}\, {\cal G}(D;\sqrt{-3{\rm i}\xi/2})\ .}
This constitutes the continuous canonical two-point function: the
quantity $\rho(D)dD$ measures the infinitesimal probability,
in the ensemble of large doubly-pointed quadrangulations,
that the two marked vertices be at a rescaled distance in the 
range $[D,D+dD]$. Its value \conttwo\ is expected to
be universal for a large class of maps forming the universality class
of so-called pure gravity. Finally, another quantity of interest is the 
(cumulative) distribution function 
\eqn\continttwo{\Phi(D)= 
{2\over {\rm i} \sqrt{\pi}} \int_{-\infty}^{\infty} d\xi\ \xi\, e^{-\xi^2}\,
{\cal F}(D;\sqrt{-3{\rm i}\xi/2})}
giving the probability that the distance is less than $D$.
The functions $\rho(D)$ and $\Phi(D)$ are represented in Fig.~\rhophi.
For small $D$, we have $\rho(D)\sim (3/7) D^3$.

\newsec{The three-point function}

We now come to our main subject, namely the three-point function 
of quadrangulations.  In this section, we present our main result, 
which is an explicit formula for the generating function
of planar quadrangulations with three marked vertices at prescribed pairwise 
distances. We shall concentrate here on the properties of this three-point
function, leaving its precise 
derivation to sections 3 and 4 below. An important application is
the continuum scaling limit in which the three-point function takes a
particularly simple universal form.  

\subsec{The three-point function of quadrangulations}

We consider planar quadrangulations with {\it three distinct vertices} 
distinguished as $1$, $2$ and $3$, and we denote by $d_{12}$, $d_{23}$ and 
$d_{31}$ their
respective pairwise graph distances. Prescribing these distances, let 
$G(d_{12},d_{23},d_{31};g)$ denote the generating function for the number 
of such triply-pointed quadrangulations, counted with a 
weight $g$ per face. Note that there are no symmetry factors for
triply-pointed quadrangulations as the latter have no symmetries.
The function $G(d_{12},d_{23},d_{31};g)$ is defined for strictly positive 
integer values of $d_{12}$, $d_{23}$ and $d_{31}$ that furthermore satisfy:
\item{-} the triangular inequalities: $d_{12} \leq d_{23}+d_{31}$ and its 
cyclic permutations;
\item{-} the condition that $d_{12}+d_{23}+d_{31}$ is even.
\par
\noindent This latter condition stems from the fact that every planar
quadrangulation is bipartite, hence the length of any cycle on the map
is even. 
These conditions are best expressed through the parametrization 
\eqn\param{\eqalign{
d_{12}&=s+t\ , \cr
d_{23}&=t+u\ ,  \cr
d_{31}&=u+s\ , \cr}}
easily inverted into 
\eqn\invpara{\eqalign{
s &= {d_{12}-d_{23}+d_{31} \over 2}\ , \cr
t &= {d_{12}+d_{23}-d_{31} \over 2}\ , \cr
u &= {-d_{12}+d_{23}+d_{31} \over 2}\ . \cr
}}
\fig{A schematic picture of the relations \param\ and 
\invpara.}{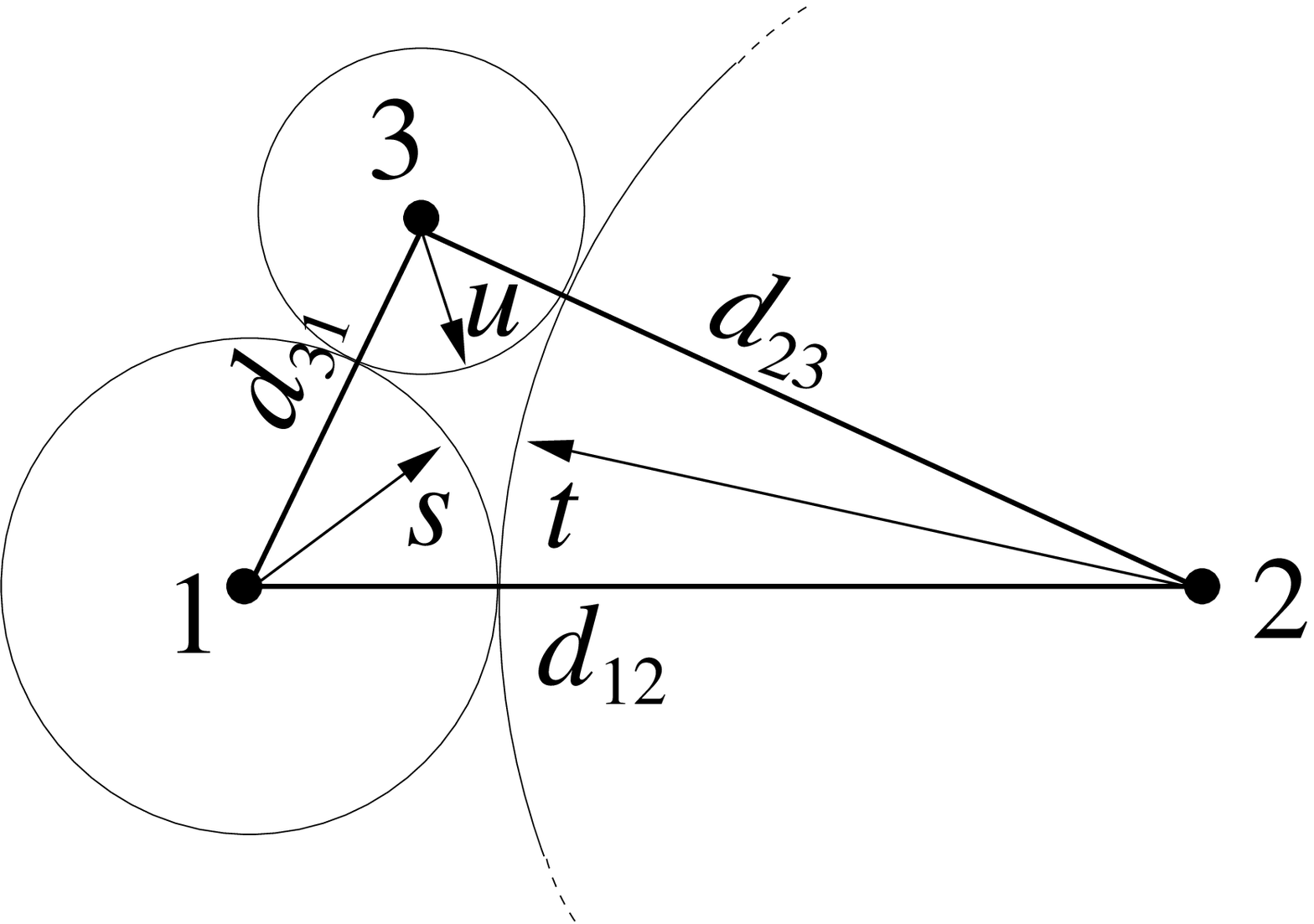}{8.cm}
\figlabel\triangle
\noindent With this parametrization, the above conditions are equivalent 
to requiring that $s$, $t$, $u$ be non-negative integers and that at most 
one of them may vanish. 
An intuitive picture of this parametrization is as 
follows (see Fig~\triangle): drawing a triangle 
$123$ in the plane with sides $d_{12}$, $d_{23}$ and $d_{31}$, the quantities
$s$, $t$ and $u$ can be viewed as the radii of three pairwise tangent circles
centered at the points $1$, $2$, $3$.
Note that having one of the parameters vanishing, say $u$, corresponds in 
this picture to having $3$ lying on the segment $12$. On the quadrangulation, 
this simply means that $3$ lies precisely on a geodesic path between $1$ 
and $2$, so that $d_{12}=d_{23}+d_{31}$. 
We shall say in this case that the three points are {\it aligned}
and that $3$ is a {\it geodesic point} between $1$ and $2$.

We find the following expression for the three-point function:
\eqn\disctrois{G(d_{12},d_{23},d_{31};g)=\Delta_s \Delta_t 
\Delta_u F(s,t,u;g)\ ,}
where $\Delta_s$ denotes the finite difference operator acting on an arbitrary
function $f(s)$ as:
\eqn\finitediff{\Delta_s\, f(s)\equiv f(s)-f(s-1)\ ,}
and similarly for $\Delta_t$ and $\Delta_u$, while $F(s,t,u;g)$ has the
explicit form:
\eqn\discrformu{\eqalign{&F(s,t,u;g) = \cr & \qquad {\q{3} \left(
\q{s+1} \q{t+1} \q{u+1} \q{s+t+u+3} \right)^2 \over \q{1}^3\q{s+t+1}
\q{s+t+3} \q{t+u+1} \q{t+u+3} \q{u+s+1} \q{u+s+3} } \cr}}
As before, we use the notation \defq\ with a parameter $x$ related 
to $g$ by Eq.~\xRtog, or more explicitly by Eq.~\Rxexplicit.
Note that $F(s,t,u;g)$ is manifestly symmetric in $s$, $t$, $u$, which is
consistent with $G$ being symmetric in the distances, as expected.

Relations \disctrois\ and \discrformu\ hold a priori only in the allowed 
range of distances,
namely when $s$, $t$, $u$ are non-negative integers and no more than 
one of them may vanish. In the case of aligned points, say when $u$ vanishes, 
relation \disctrois\ reduces to 
\eqn\aligned{G(s+t,t,s;g)=\Delta_s \Delta_t F(s,t,0;g)}
as $F(s,t,-1;g)$ vanishes identically, while relation \discrformu\ reduces to
\eqn\fstzero{F(s,t,0;g)={\q{3}\, \q{s+1}\, \q{t+1}\, \q{s+t+3} 
\over \q{1}\, \q{s+3}\, \q{t+3}\, \q{s+t+1}}\ .}
It is convenient to slightly extend the domain of definition of $G$
and $F$ as follows:
when exactly two of the parameters, say $t$ and $u$, vanish, 
corresponding to 2 and 3 being identical, Eqs.~\disctrois\ and 
\discrformu\ give a result $G(s,0,s;g)=0$, which can 
still be considered as the
correct result as we imposed that the three marked vertices have to be distinct.
Finally, when $s=t=u=0$, the above formulas yield the somewhat conventional 
value $G(0,0,0;g)=1$.
We shall therefore take the convention that $G(d,d,0;g)=G(d,0,d;g)=G(0,d,d;g)
=\delta_{d,0}$. With this convention, Eq.~\disctrois\ may be inverted into:
\eqn\invrel{F(s,t,u;g)= \sum_{s'=0}^s \sum_{t'=0}^t \sum_{u'=0}^u
G(s'+t',t'+u',u'+s';g)\ ,}
so that $F(s,t,u;g)-1$ can be interpreted as the generating function of 
quadrangulations with three {\it distinct} marked vertices at distances 
$d_{12}=s'+t'$, $d_{23}=t'+u'$ and $d_{31}=u'+s'$ such that $s'\leq s$, 
$t'\leq t$ and $u'\leq u$.
In particular, taking the limit $s,t,u \to \infty$ amounts to relaxing
these constraints, i.e. considering three arbitrary distinct vertices. 
We have: 
\eqn\limF{\eqalign{\lim_{s,t,u\to \infty} F(s,t,u;g)& = 
{1+x+x^2 \over (1-x)^2}  = {1\over 2}\left(1+(1-12g)^{-1/2}\right) \cr
& =1+\sum_{n\geq 1} 
{3^n \over 2} {2n \choose n} g^n \ ,\cr}}
where we identify the coefficient of $g^n$ as the number of triply-pointed
quadrangulations with $n$ faces and three distinct marked vertices.

\subsec{Universal continuum limit}

The above formula for the three-point function is easily translated in
the continuum limit, i.e. when $g$ approaches its critical value $1/12$
and all distances are large, with the following scaling:
\eqn\scal{\eqalign{g&={1\over 12}\left(1- \Lambda\, \epsilon\right) \ ,\cr
d_{12} =D_{12} \epsilon^{-1/4}\ ,\quad  d_{23}& =D_{23} \epsilon^{-1/4}\ ,
\quad d_{31}=D_{31} \epsilon^{-1/4}\ ,\cr}}
with $\epsilon\to 0$. 
Similarly, the parameters $s$, $t$, $u$ defined
in Eq.~\param\ have the scaling:
\eqn\scalpara{ s= S \epsilon^{-1/4},\ t= T \epsilon^{-1/4},\ 
s= U \epsilon^{-1/4},} 
where $S$, $T$, $U$ are related to $D_{12}$, $D_{23}$ and $D_{31}$ 
via relations similar to Eqs.~\param\ and \invpara, namely
\eqn\paramcap{\eqalign{
D_{12}&=S+T\ ,\qquad\qquad
S = {D_{12}-D_{23}+D_{31} \over 2}\ , \cr
D_{23}&=T+U\ , \qquad\qquad
T = {D_{12}+D_{23}-D_{31} \over 2}\ , \cr
D_{31}&=U+S\ ,\qquad\qquad
U = {-D_{12}+D_{23}+D_{31} \over 2}\ . \cr
}}
In this limit, we obtain 
\eqn\scalGF{\eqalign{& G(d_{12},d_{23},d_{31};g)\sim 
\epsilon^{1/4} \, 2{\cal G}(D_{12}, D_{23},D_{31};\sqrt{3/2}\Lambda^{1/4}) \cr
& F(s,t,u;g)\sim \epsilon^{-1/2} {\cal F}(S,T,U;\sqrt{3/2}\Lambda^{1/4}) 
\cr}}
with finite scaling functions ${\cal G}$ and ${\cal F}$ depending on the
rescaled distances and on the ``cosmological constant'' $\Lambda$, encoded
for later convenience in the parameter
\eqn\valalp{\alpha=\sqrt{3/2}\Lambda^{1/4}\ .}
Note the factor of $2$ in the first line of Eq.~\scalGF\ which we introduced
to compensate the fact that this relation holds only for discrete distances 
satisfying the parity condition that $d_{12}+d_{23}+d_{31}$ be even, while
$G$ should be taken as zero otherwise. The factor $2$ ensures a posteriori 
that ${\cal G}=(2{\cal G}+0)/2$ 
is a correct measure of $G$ on average.
Relation \disctrois\ becomes
\eqn\conttrois{{\cal G}(D_{12}, D_{23},D_{31};\alpha) = {1\over 2}
\partial_S \partial_T \partial_U {\cal F}(S,T,U;\alpha)\ ,}
while Eq.\discrformu\ translates immediately into
\eqn\gdcont{{\cal F}(S,T,U;\alpha) = {3\over \alpha^2}
\left({\sinh(\alpha(S+T+U)) \sinh(\alpha S) \sinh(\alpha T) \sinh(\alpha U) 
\over \sinh(\alpha(S+T)) \sinh(\alpha(T+U)) \sinh(\alpha(U+S))
}\right)^2 \ .}
The above formulas \conttrois\ and \gdcont\ give the continuous three-point 
function in the grand-canonical ensemble with a varying size of the 
quadrangulations and a fixed cosmological constant $\Lambda$. 
This constitutes the grand-canonical three-point function of
pure 2D quantum gravity.

If we wish instead to work in the canonical ensemble
where the number of faces $n$ of the quadrangulations is fixed to a large
value, and the distances are scaled as $n^{1/4}$, we can use the above
formulas with $\epsilon=1/n$, and a varying cosmological constant 
$\Lambda=-\xi^2$. As seen for the two-point function, the 
coefficient of $g^n$ in $G$ or $F$
is obtained by a contour integral in $g$ which, asymptotically at large $n$, 
yields, via a saddle point estimate, an integral over real values of $\xi$, 
namely: 
\eqn\canon{\eqalign{
G(d_{12},d_{23},d_{31};g)\vert_{g^n} & \sim 
{12^n \over {\rm i} \pi n} 
\int_{-\infty}^{\infty} d\xi\ \xi\, e^{-\xi^2}\, n^{-1/4}
\, 2{\cal G}(D_{12},D_{23},D_{31};\sqrt{-3{\rm i}\xi/2})\ ,\cr
F(s,t,u;g)\vert_{g^n} & \sim 
{12^n \over {\rm i} \pi n} 
\int_{-\infty}^{\infty} d\xi\ \xi\, e^{-\xi^2}\, n^{1/2}
{\cal F}(S,T,U;\sqrt{-3{\rm i}\xi/2})\ ,\cr }}
where we take the determination $\sqrt{-{\rm i}\tau}= 
e^{-{\rm sign}(\tau){\rm i}\pi/4} \sqrt{|\tau|}$ for $\tau$ real.
Normalizing these asymptotic values, we find the canonical three-point function
\eqn\contp{\rho(D_{12},D_{23},D_{31})= 
{2\over {\rm i} \sqrt{\pi}} \int_{-\infty}^{\infty} d\xi\ \xi\, e^{-\xi^2}\,
{\cal G}(D_{12},D_{23},D_{31};\sqrt{-3{\rm i}\xi/2})\ ,}
where $\rho(D_{12},D_{23},D_{31})\, dD_{12}\, dD_{23}\, dD_{31}$ {\it is the 
infinitesimal probability that the pairwise rescaled distances between 
the three marked points in the ensemble of triply-pointed random 
quadrangulations of fixed large size be respectively in the ranges} 
$[D_{12},D_{12}+dD_{12}]$, $[D_{23},D_{23}+dD_{23}]$, $[D_{31},D_{31}+dD_{31}]$.
Eq.~\contp\ constitutes the canonical three-point function of pure 2D 
quantum gravity for the planar topology.

Eq.~\contp\ was obtained by dividing the first line of Eq.~\canon\ by 
the number of triply pointed quadrangulations of size $n$, as given by \limF:
\eqn\tpq{{3^n\over 2} {2n\choose n} \sim {12^n\over 2 \sqrt{\pi} n^{1/2}}\ ,}
and multiplying by $n^{3/4}/2$ which is the number of allowed triplets of 
discrete distances $(d_{12},d_{23},d_{31})$ falling into the above range of 
rescaled distances (note the factor $1/2$ coming from the parity of
$d_{12}+d_{23}+d_{31}$).
Similarly, the second line of Eq.~\canon\ translates into the integrated
three-point function 
\eqn\contpint{\Phi(S,T,U)= 
{2\over {\rm i} \sqrt{\pi}} \int_{-\infty}^{\infty} d\xi\ \xi\, e^{-\xi^2}\,
{\cal F}(S,T,U;\sqrt{-3{\rm i}\xi/2})\ ,}
corresponding to the {\it probability that the three marked points be at
distances $S'+T'$, $T'+U'$ and $U'+S'$ with 
$S'\leq S$, $T'\leq T$ and $U'\leq U$}. The functions $\Phi$ and $\rho$ 
are related through
\eqn\phitorho{\rho(D_{12},D_{23},D_{31})= {1\over 2}
\partial_S \partial_T \partial_U \Phi(S,T,U)}
or conversely
\eqn\rhotophi{\Phi(S,T,U)= \int_0^S \!dS' \int_0^T \!dT'\int_0^U \!dU'\ 2 
\rho(S'+T',T'+U', U'+S')\ ,}
where the factor $2$ may be alternatively understood as the Jacobian
in the change of variables from $(D_{12},D_{23},D_{31})$ to $(S,T,U)$.
Finally, Eq.~\limF\ translates into $\lim_{S,T,U\to \infty} \Phi(S,T,U)=1$, 
so that we find the normalization 
\eqn\normrho{\int_{\cal D}dD_{12}\, dD_{23}\, dD_{31}\, 
\rho(D_{12}, D_{23},D_{31}) = 1}
as it should. Here ${\cal D}$ denotes the domain of positive real values
of $D_{12}$, $D_{23}$ and $D_{31}$ satisfying the triangular inequalities.
 
Let us know discuss in more details the properties of the above continuous
grand-canonical or canonical three-point functions. As a preliminary exercise, 
let us see how to recover the two-point function 
from the value of the three-point function upon integrating over 
the position of one of the points, say 3, 
keeping the distance $D_{12}$ fixed to a constant value. We have
\eqn\intthree{\eqalign{\int_{{\cal D}_{12}} dD_{23}\, dD_{31}\, 
&{\cal G}(D_{12}, D_{23},D_{31};\alpha)\cr & = 
\int_0^\infty \!dS \int_0^\infty \!dT \int_0^\infty \!dU \ 
\partial_S\partial_T\partial_U {\cal F}(S,T,U;\alpha)\, \delta(S+T-D_{12}) \cr
&=\int_0^\infty \!dS \int_0^\infty \!dT \ \lim_{U\to \infty}\left(
\partial_S\partial_T {\cal F}(S,T,U;\alpha)\right)\, \delta(S+T-D_{12})  \cr
&= \int_0^{D_{12}} \!dS\ 18 \left({\sinh(\alpha S) \sinh(\alpha (D_{12}-S))\over
\sinh^2(\alpha D_{12})}\right)^2 \cr 
& = {9\over 8 \sinh^4(\alpha D_{12})} \left(4D_{12} +2D_{12} \cosh(2 \alpha 
D_{12})
-3\, {\sinh(2\alpha D_{12})\over \alpha}\right) \ .\cr}}
Here ${\cal D}_{12}$ is a short-hand notation for the domain of
allowed distances $D_{23}$, $D_{31}$ at a fixed value of $D_{12}$.
Comparing this result with the value \valGtwo\ for the
continuous two-point function, we see that
\eqn\compthreetwo{\int_{{\cal D}_{12}} dD_{23}\, dD_{31}\, 
{\cal G}(D_{12}, D_{23},D_{31};\alpha)= 
-{9\over 16 \alpha^3} \partial_\alpha {\cal G}(D_{12};\alpha)\ .}
Taking $\alpha$ as in Eq.~\valalp\ for the grand-canonical ensemble, 
we have $-9/(16 \alpha^3) \partial_\alpha= - \partial_\Lambda$ 
and we thus find the expected result that the integral
of the grand-canonical three-point function over the position
of one of the points reproduces precisely the grand-canonical
two-point function for the remaining two points, up to the action of 
the trivial operator $-\partial_\Lambda$ corresponding precisely to the 
marking of the third point with no prescription on its position.

Similarly in the canonical ensemble, taking $\alpha=\sqrt{-3{\rm i}\xi/2}$,
we have $-9/(16 \alpha^3) \partial_\alpha= 1/(2\xi)\partial_\xi$
so that
\eqn\sumrulecont{\int_{{\cal D}_{12}} dD_{23} \, dD_{31}\,
\rho(D_{12}, D_{23},D_{31}) = 
{2\over {\rm i} \sqrt{\pi}} \int_{-\infty}^{\infty} d\xi\ \xi\, e^{-\xi^2}\,
{1\over 2\xi}\partial_{\xi} {\cal G}(D_{12};\sqrt{-3{\rm i} \xi /2}) 
= \rho(D_{12})}
upon integrating by part over $\xi$, with $\rho(D_{12})$ as in Eq.~\conttwo. 
As expected, we recover the canonical two-point function as the marginal
probability density obtained by integrating the canonical three-point function
over the position of one of the points.
\fig{Left: plots of the conditional probability density 
$\rho(D_{23},D_{31}\vert D_{12})$ for $D_{12}=0.8$, $1.5$ and $3.0$, from
top to bottom. Right: the corresponding contour plots, where 
the thick lines indicate the boundary of the allowed domain 
${\cal D}_{12}$ for distances, as determined by the triangular inequalities.
For the first (small enough) two values of $D_{12}$, the density is maximal 
for $D_{23}=D_{31}\sim 1.5$. For the larger value $D_{12}=3.0$, the
density becomes elongated and squeezed along the boundary 
line $D_{23}+D_{31}=D_{12}$.}{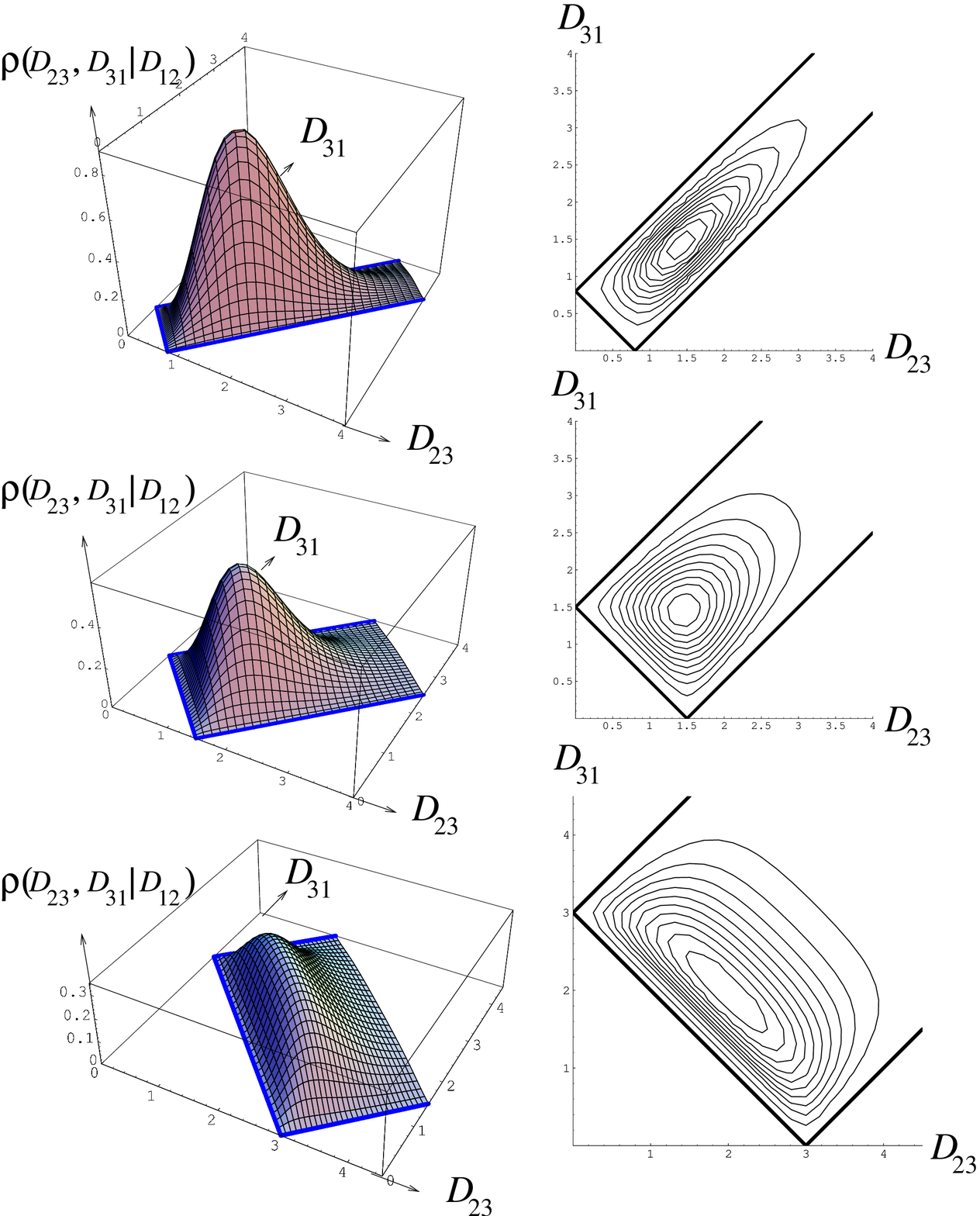}{12.cm}
\figlabel\plots
In order to visualize the three-point function and in view of
the above result, it is convenient
to first fix one of the distances, say $D_{12}$ to a constant value 
and to plot the {\it conditional probability density}: 
\eqn\rhocond{\rho(D_{23},D_{31}\vert D_{12})\equiv {\rho(D_{12},D_{23},D_{31}) \over
\rho(D_{12})}\ .}
The quantity $\rho(D_{23},D_{31}\vert D_{12}) dD_{23} dD_{31}$ 
measures the infinitesimal conditional probability
that point $3$ be at distances from points $1$ and $2$ respectively in the 
ranges $[D_{23},D_{23}+dD_{23}]$ and $[D_{31},D_{31}+dD_{31}]$, given that 
the distance between $1$ and $2$ is $D_{12}$.
This conditional probability density may then be plotted 
for various values of $D_{12}$,
as illustrated in Fig.~\plots, where a number of properties of the three-point 
function can be observed.

First, the three-point function vanishes at the boundary of the domain
${\cal D}_{12}$, namely when the three points are aligned. It also vanishes
when the distances from point $3$ to points $1$ and $2$ tends to infinity. 
The probability density is maximal when $D_{23}$ and $D_{31}$ are both 
equal to a particular value $D_{\rm max}$, depending on $D_{12}$. For small 
enough $D_{12}$ (less than $\sim 3$ or so), this preferred value is roughly
independent of $D_{12}$ and comparable to the maximum of the two-point function 
$\rho(D)$, namely $D_{\rm max}\sim 1.5$. For larger $D_{12}$, the triangular 
inequalities impose that $D_{\rm max}\ge D_{12}/2$ so that the value of 
$D_{\rm max}$ must increase. In this case, the density profile becomes 
squeezed and elongated along the boundary line $D_{23}+D_{31}=D_{12}$ of the
domain ${\cal D}_{12}$ with, as we shall see below, a width of order 
$D_{12}^{-1/3}$ at large $D_{12}$.

\subsec{Limiting behaviors}
\fig{The conditional probability density $\rho(D_{23},D_{31}\vert D_{12})$
for a small value of $D_{12}$, here $D_{12}=0.05$ (a). The same density
(b) and its contour plot (c), now expressed as a function of the 
longitudinal variable $U=(D_{23}+D_{31}-D_{12})/2$ 
and a {\it rescaled} transverse variable $\omega=(D_{31}-D_{23})/D_{12}$.
As apparent by taking longitudinal and transverse cut views 
along the thick lines in (c), the conditional probability 
factorizes into the product of 
the two-point function $\rho(U)$ (red curve in (d)) and the probability density
$\psi(\omega)$ (green curve in (e)).}{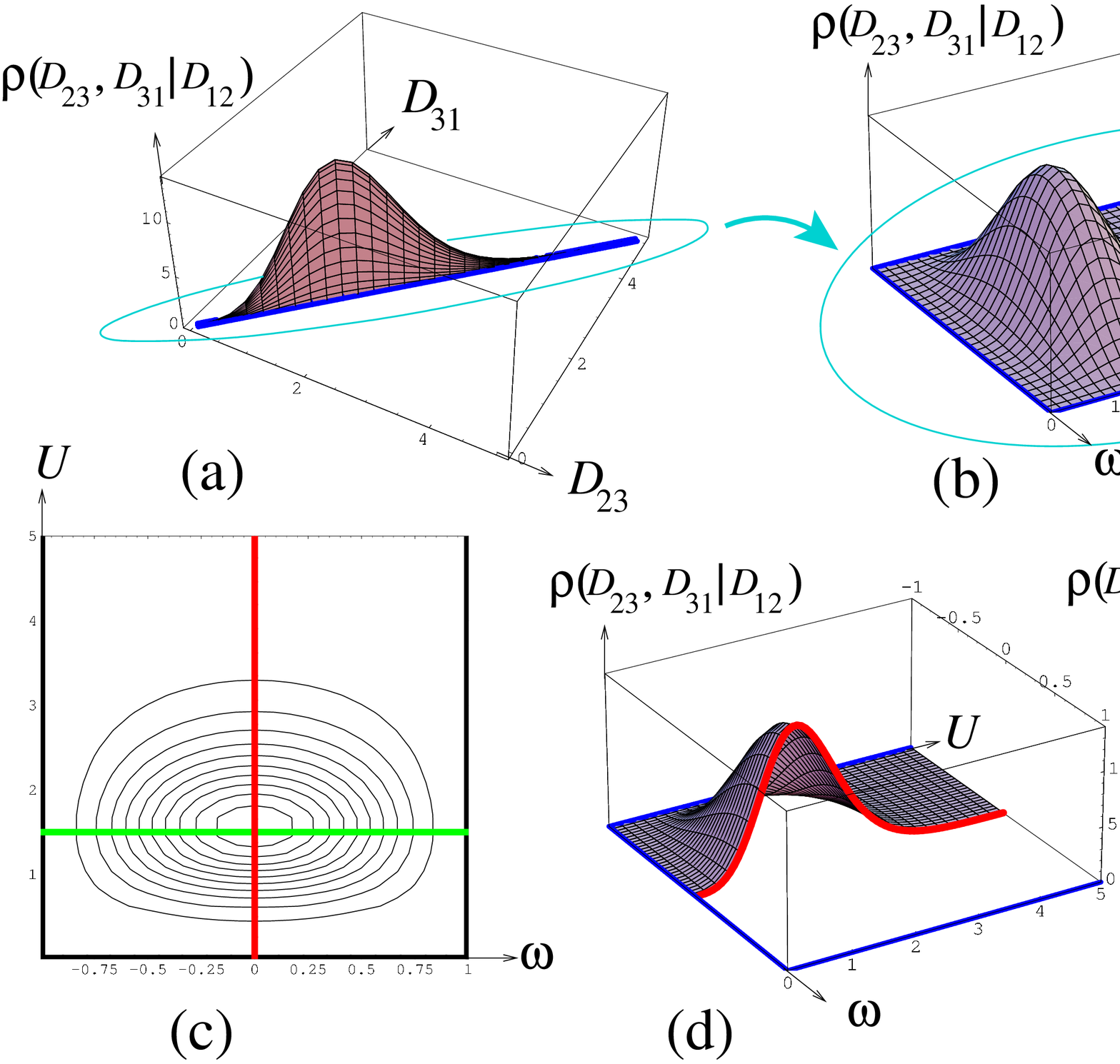}{14.cm}
\figlabel\dsmall
The conditional probability density 
$\rho(D_{23},D_{31}\vert D_{12})$ has interesting 
limiting behaviors whenever $D_{12}$ becomes small or large. For $D_{12}\to 0$,
taking $S=\sigma D_{12}$ and $T=\tau D_{12}$, we find that
\eqn\zerolim{\eqalign{{\cal G}\big((\sigma\!+\!\tau)D_{12},\tau D_{12}+U,
\sigma D_{12}+U & ;\alpha
\big)
={1\over 2 D_{12}^2}
\partial_\sigma\partial_\tau\partial_U{\cal F}(\sigma D_{12},\tau D_{12}, 
U;\alpha) \cr
&= {9 D_{12}^2\over 2}\  {\sigma^2 \tau^2 (\sigma^2+4\sigma \tau +\tau^2)
\over (\sigma+\tau)^4} {\cal G}(U;\alpha) +{\cal O}(D_{12}^3)\ ,\cr}}
where ${\cal G}(U;\alpha)$ is the two-point function, as given by
Eq.~\valGtwo. Taking $\sigma+\tau=1$ and using the small $D_{12}$ 
behavior $\rho(D_{12})\sim (3/7) D_{12}^3$, we deduce that
\eqn\zerorho{\rho\big((1-\sigma) D_{12}+U,\sigma D_{12}+U\vert D_{12}\big) 
\sim {1\over 2 D_{12}} \times 
\rho(U) \times 21 
\sigma^2 (1-\sigma)^2 \big(1+2\sigma(1-\sigma)\big)
}
in terms of the canonical two-point function $\rho(U)$. Writing 
$U=(D_{23}+D_{31}-D_{12})/2$ and $\sigma=(1+\omega)/2$ with 
$\omega=(D_{31}-D_{23})/D_{12}$, this equation may alternatively be
written as 
\eqn\zerorhobis{\eqalign{&\rho(D_{23},D_{31}\vert D_{12})\sim {1\over D_{12}}
\times \rho\left(
{D_{23}+D_{31}-D_{12} \over 2}\right) \times \psi\left({D_{31}-D_{23} \over
 D_{12}}\right)
\cr &{\rm with}\quad \psi(\omega)\equiv {21\over 64} (1-\omega^2)^2 (3-\omega^2)
\cr}}
involving a new function $\psi(\omega)$ properly normalized to $1$ when
$\omega$ varies from $-1$ to $1$. The conditional probability density
$\rho(D_{23},D_{31}\vert D_{12})$ therefore factorizes in the small 
$D_{12}$ limit into the product of the canonical two-point function in the 
``longitudinal direction'' (corresponding to varying values of $D_{23}+D_{31}$)
and of the above probability density 
$\psi$ in the ``transverse direction'' (corresponding
to varying $D_{31}-D_{23}$). This property is illustrated in Fig.~\dsmall\
for a value $D_{12}=0.05$.
\fig{The conditional probability density $\rho(D_{23},D_{31}\vert D_{12})$
for a large value of $D_{12}$, here $D_{12}=10.$ (a). The same density
(b) and its contour plot (c), now expressed as a function of the 
transverse variable $S=(D_{12}+D_{31}-D_{23})/2$ 
and a {\it rescaled} longitudinal variable 
$\nu=(9D_{12})^{1/3}(D_{23}+D_{31}-D_{12})/2$.
As apparent by taking longitudinal and transverse cut views 
along the thick lines in (c), the conditional probability 
factorizes into the product of 
the probability density $\varphi(\nu)$ (red curve in (d)) 
and a uniform density in the $S$ direction (green curve in 
(e)).}{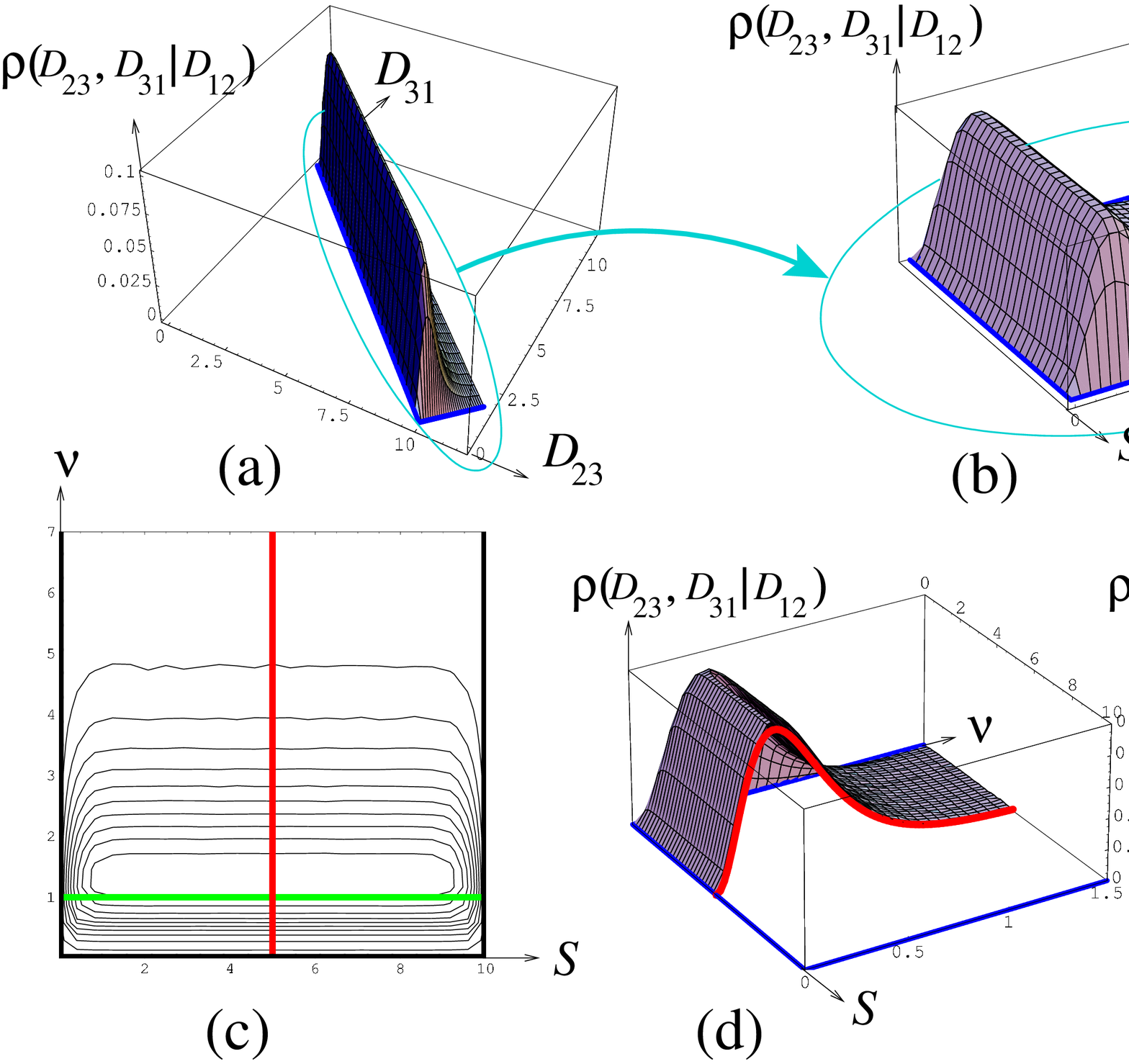}{14.cm}
\figlabel\dlarge
Let us now consider the case $D_{12}\to \infty$, and more precisely the 
limit when {\it both} $S$ and $T$ tend simultaneously to infinity 
(with $D_{12}=S+T$).
In this limit, we have
\eqn\larD{{\cal G}(D_{12},(D_{12}-S)+U,S+U;\alpha)\!\sim\!
24\alpha
e^{-\alpha (2D_{12}+5U)}\sinh^2(\alpha U)\left(3
\cosh(\alpha U)\!+\!19\sinh(\alpha U)\right)
}
while the grand-canonical two-point function behaves as
\eqn\asymgc{{\cal G}(D_{12};\alpha) \sim 16 \alpha^3 \, e^{- 2 \alpha D_{12}}
\ .}
We deduce that
\eqn\rhoasy{\eqalign{\rho(D_{12}&-S+U,S+U \vert D_{12}) \cr &\sim 
{
\int_{-\infty}^{\infty} d\xi\,\xi e^{-\xi^2}
24\alpha
\, e^{-\alpha (2D_{12}+5U)}\, \sinh^2(\alpha U)\left(3
\cosh(\alpha U) + 19\sinh(\alpha U)\right)
\over 
\int_{-\infty}^{\infty} d\xi\,\xi e^{-\xi^2}
16 \alpha^3 \, e^{- 2 \alpha D_{12}}
}\ ,\cr}}
with $\alpha=\sqrt{-3 {\rm i}\xi/2}$ . Note that this result is independent
of $S$. Both integrals are dominated
at large $D_{12}$ by their saddle point at $\xi=(3 D_{12}^2)^{1/3} {\rm i}/2$,
leading to 
\eqn\rhoasybis{\eqalign{&\rho(D_{23},D_{31} \vert D_{12}) 
\sim {1\over 2 D_{12}}\times (9 D_{12})^{1/3}
\ \varphi\left((9D_{12})^{1/3}\, {D_{23}+D_{31}-D_{12} \over 2}\right) \cr
&{\rm with}\quad  \varphi(\nu)\equiv 
{4\over 3} \sinh(\nu/2)^2 \left(11 e^{-2 \nu}-8 e^{-3 \nu}\right)\cr}}
involving a new function $\varphi(\nu)$ properly normalized to $1$ when
$\nu$ varies from $0$ to $\infty$. In the limit of 
large $D_{12}$, the conditional probability density
$\rho(D_{23},D_{31}\vert D_{12})$ therefore becomes elongated and squeezed
along the line $D_{23}+D_{31}=D_{12}$, with a profile given by 
the above probability density $\varphi$ in the
longitudinal direction and a uniform probability density
in the transverse direction. This property is illustrated in Fig.~\dlarge\
for a value $D_{12}=10$. Note that the extension in the longitudinal
direction is of order $D_{12}^{-1/3}$, as announced.

To end this section, let us finally discuss the value of the three-point
function whenever all distances become small or large. For
$S$, $T$ and $U$ small (and of the same order), we find the limiting behavior
\eqn\Phismall{\Phi(S,T,U) \sim {9\over 28}\, {\left( S T U (S+T+U)\right)^3
(S^2+T^2+U^2+S T+T U +U S)\over (S+T)^2 (T+U)^2 (U+S)^2 }\ ,}
which is a homogeneous function of degree $8$ of its arguments.
Applying Eq.~\phitorho, we deduce that $\rho(D_{12},D_{23},D_{31})$
is, at small distances, a homogeneous function of degree $5$ of
its arguments. Note that for a manifold of fractal dimension $d_F$, 
we expect a degree $(d_F-1)+(d_F-2)=2 d_F-3$ as the choice of point $2$ at
a fixed distance from point $1$ will select a manifold of dimension
$d_F-1$ and the choice of point 3 at fixed distances from points 1 and 2
will then select a manifold of dimension $d_F-2$. 
The value $5$ above is therefore compatible with the known value 
$d_F=4$ for the fractal dimension of large quadrangulations.

Finally, when $D_{12}$, $D_{23}$ and $D_{31}$ are large and of the same order,
we have 
\eqn\gasylar{{\cal G}(D_{12},D_{23},D_{31};\alpha)\sim
66\ \alpha\ e^{-\alpha(D_{12}+D_{23}+D_{31})}}
and we find the asymptotic behavior
\eqn\rhoasylar{\rho(D_{12},D_{23},D_{31})\sim
{99\over \sqrt{6}} (D_{12}+D_{23}+D_{31}) e^{-\left({3\over 4}\right)^{5/3}
(D_{12}+D_{23}+D_{31})^{4/3}}\ .}

\newsec{Triply-pointed quadrangulations and well-labeled maps}

Let us now come to the derivation of our main result, as
given by Eqs.~\disctrois\ and \discrformu. For this derivation,
we shall rely on a new bijection discovered recently by Miermont
between multiply-pointed quadrangulations, i.e. quadrangulations
with a number, say $p$, of marked vertices, and {\it well-labeled
maps} with $p$ faces, as will be defined below. This bijection
extends a well-known bijection obtained by Schaeffer [\xref\MS,\xref\CS]
between pointed quadrangulations and well-labeled trees (or g-trees)
corresponding to the case $p=1$. In this section, we first recall 
the Miermont bijection and show how to use it to treat the case of 
planar triply-pointed quadrangulations with prescribed pairwise 
distances between the three marked vertices. We then exploit these 
results in section 4 to obtain explicit formulas for various generating 
functions, leading eventually to expressions \disctrois\ and \discrformu.

\subsec{The Miermont bijection}

The Miermont bijection works as follows. We start with a {\it
bipartite} quadrangulation of genus $h$ with $n$ faces, together with
a marked $p$-tuple of {\it distinct} vertices, hereafter denoted
$1,2,\ldots,p$ and referred to as the {\it sources} of the
quadrangulation. We shall denote by $d_{ij}$, $i,j=1,\ldots,p$, the
distance in the quadrangulation between the sources $i$ and $j$. For
the construction below to work, we must impose that $d_{ij}\geq 2$ for
all $i\neq j$, i.e. no two marked vertices are immediate neighbors on
the map.

With each source $i$, we furthermore associate an integer $\tau_i$, hereafter 
referred to as the {\it delay} of the source $i$, and which will act as 
a penalty when measuring distances from this source. Delays are subject 
to the following constraints:
\eqn\contdelay{\eqalign{& \vert \tau_i-\tau_j\vert < d_{ij},\quad  
1 \le i\neq j \leq p \ ,\cr & \tau_i-\tau_j+d_{ij} \ \hbox{is even,}\quad
1\leq i,j\leq p\ .\cr}}
Note that delays satisfying these constraints can always be found, 
for instance by taking $\tau_i=0$ (respectively $1$) for all sources
belonging to the white (respectively black) sublattice of the naturally
colored bipartite quadrangulation.

Given a quadrangulation with sources and delays as above, we may
associate with each vertex $v$ of the quadrangulation a label $\ell(v)$
defined as \eqn\defell{\ell(v)=\min_{j=1,\ldots,p}(\tau_j+d_j(v))\ ,}
where $d_j(v)$ denotes the distance in the quadrangulation from the
vertex $v$ to the source $j$. The label of any vertex is therefore the
minimal value for all sources of a ``delayed distance'' equal, for
each source, to the actual graph distance to this source incremented
by the delay of the source.  The first condition in \contdelay\
ensures that the label of the source $i$ is $\tau_i$ (with the minimum
in \defell\ reached only for $j=i$). Clearly, the labels of two
adjacent vertices cannot differ by more than $1$. Moreover, the parity
of $\tau_j+d_j(v)$ necessarily changes between neighbors on a
bipartite map and, from the second condition in \contdelay, this
parity is independent of $j$. This implies that the parity of
$\ell(v)$ must also change between neighbors, leading to the crucial
property:
\fig{The two possible types of faces according to the labels of their 
incident vertices: (a) simple faces and (b) confluent faces. 
With each face, we associate an edge (thick line) as shown.}{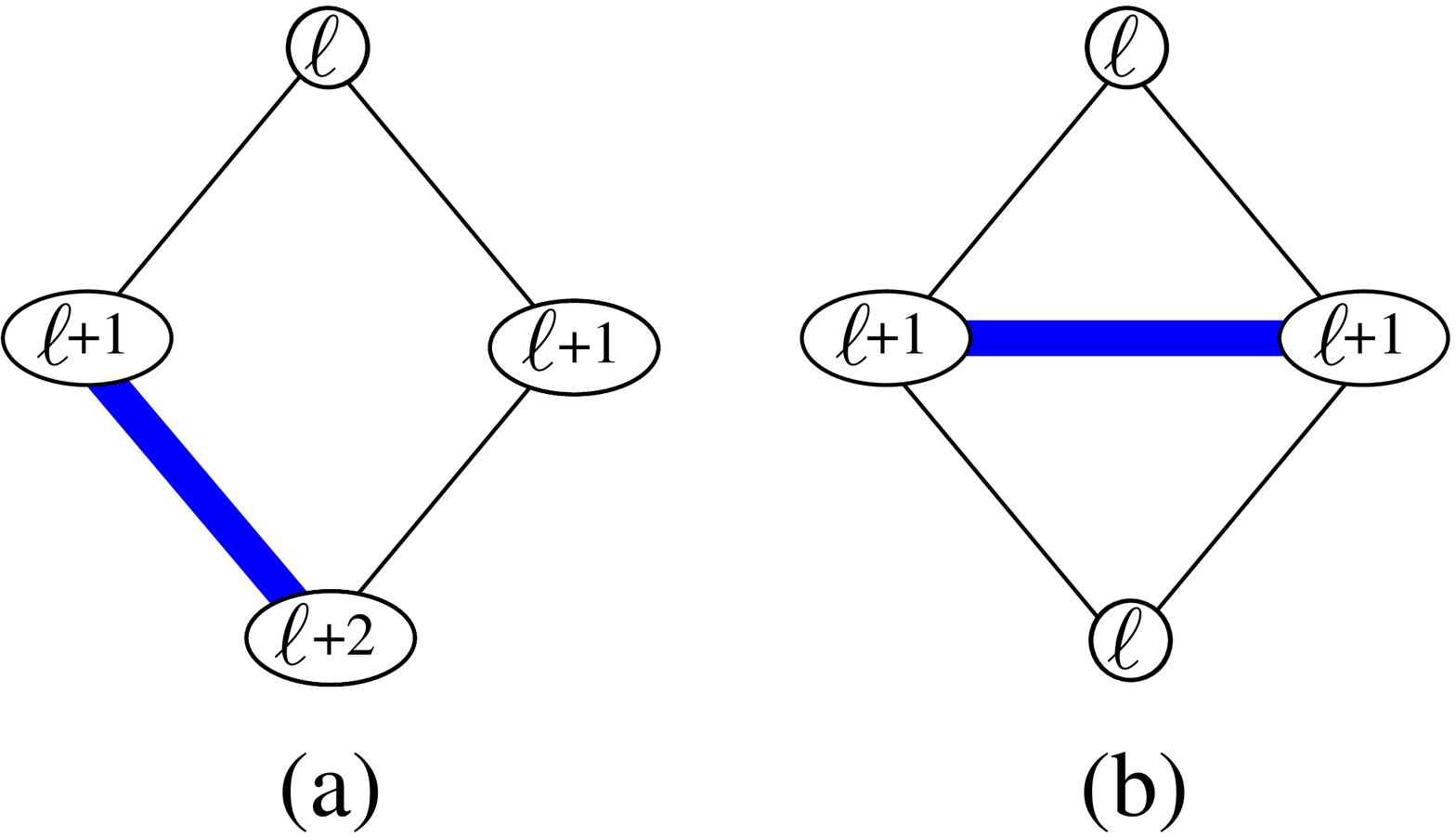}{8.cm}
\figlabel\faces
\eqn\deltal{\vert \ell(v)-\ell(v')\vert =1}
for any pair of adjacent vertices $v$ and $v'$. The faces of the 
quadrangulation are therefore of two types: 
simple faces with a cyclic sequence of labels around the face of the 
form $\ell$, $\ell+1$, $\ell+2$, $\ell+1$ and confluent faces with a 
sequence of the form $\ell$, $\ell+1$, $\ell$, $\ell+1$. As in Schaeffer's 
original construction, we associate with each of the $n$ faces of the 
(now labeled) quadrangulation an edge as follows: 
for each simple face, we select the edge 
of type $\ell+2 \to \ell+1$ encountered clockwise around the face 
(see Fig.\faces-(a)); for each confluent face, we draw a new edge 
between the two vertices labeled $\ell+1$ (see Fig.\faces-(b)). 
\fig{An example (a) of triply-pointed quadrangulations where
the marked vertices are indicated by thick circles. Taking
delays $\tau_1=0$, $\tau_2=1$ and $\tau_3=2$ as indicated, 
each vertex $v$ receives its label $\ell(v)$ as defined by \defell.
Applying (b) the rules of Fig.\faces\ on all the labeled faces
results in a well-labeled map (c) with three distinguished faces $i=1,2,3$.
As apparent in (c), the minimal label among vertices incident to 
face $i$ is $\tau_i+1$.}{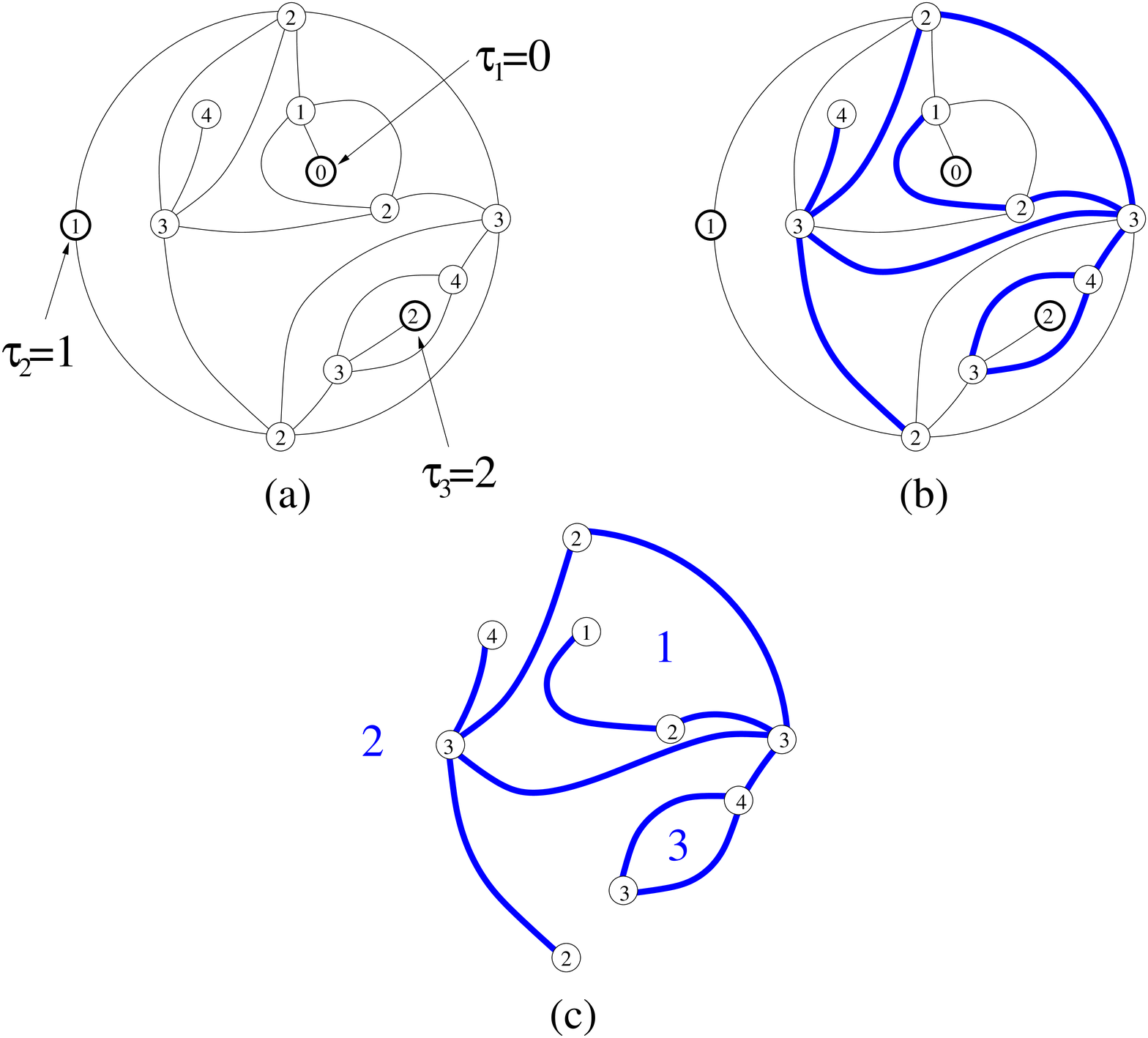}{14.cm}
\figlabel\quadtomap
It was shown by Miermont that the graph spanned by these $n$ edges
is a map that connects all the original vertices but the sources, 
which become isolated in the process (see Fig.~\quadtomap\ for an
example). This map has moreover the same genus $h$ 
as the original quadrangulation, and has $p$ faces, each of them enclosing 
exactly one of the original sources and being referred to as face 
$1,2,\ldots,p$ accordingly. Keeping the labels on the vertices of the map, 
we see that, by construction, these {\it labels vary by at most one 
along any edge of the map}. Maps with integer labels satisfying this latter
constraint will be called {\it well-labeled} maps. Finally, 
it can be shown that, for all the vertices incident to face $i$ 
in the well-labeled map, the minimum in \defell\ is reached for $j=i$ 
(and possibly for some other value of $j$). In particular,  
all these vertices have a label larger than or equal to $\tau_i+1$, 
and this value is reached by at least one of them, namely:
\eqn\conlab{\min_{v\ {\rm incident\ to\ face}\ i} \ell(v)= \tau_i+1\ .}
\medskip
It was shown that the construction above provides a bijection between,
on the one hand, quadrangulations of genus $h$ with $n$ faces, $p$ 
distinguished sources and {\it prescribed delays} $\{\tau_i\}$
satisfying \contdelay, 
and on the other hand, well-labeled maps of genus $h$ with $n$ edges and
$p$ distinguished faces whose labels satisfy \conlab.
\fig{The inverse construction leading back from a well-labeled map,
here with $p=3$ faces, to a multiply- (here triply-)pointed
quadrangulation. As explained in the text, calling $\tau_i+1$ 
the minimal label of vertices incident to face $i$, an extra vertex
(thick circle) with label $\tau_i$ is added at the center of each 
face $i$. Each corner with label $\ell$ is connected (a) by an arch 
(dashed line) to its successor with label $\ell-1$. The
set of these arches reconstructs the edges of the 
quadrangulation (b), while the added vertices form the sources
of the quadrangulation.}{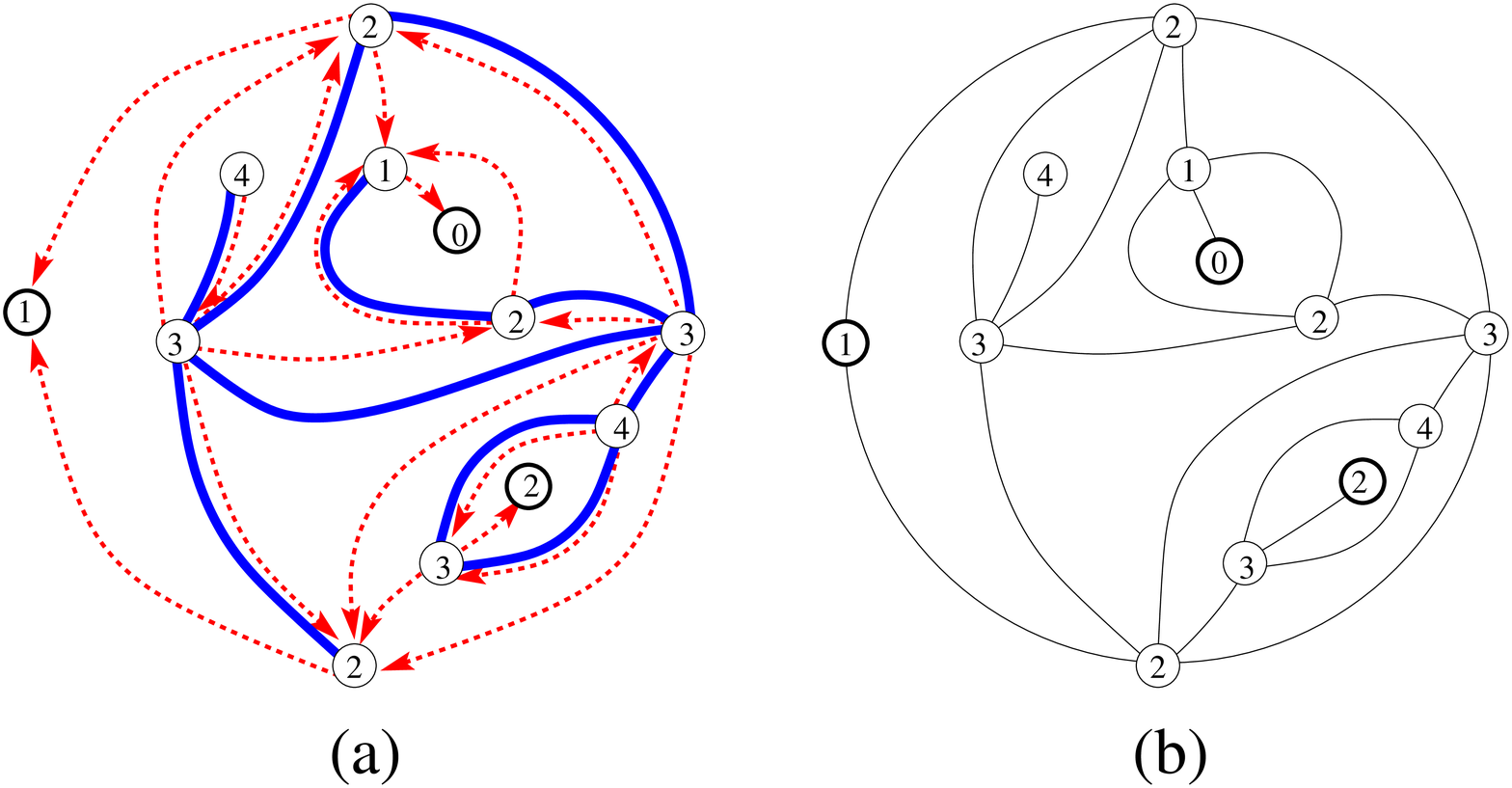}{12.cm}
\figlabel\maptoquad
To recover the original quadrangulation from the well-labeled map, 
one can proceed as follows: for each face $i$ of the map, we visit
all its incident corners successively by going counterclockwise around 
the face. Calling $\tau_i+1$ the minimal label of vertices incident
to face $i$, we then connect by an arch each corner with label 
$\ell>\tau_i+1$ to the first corner labeled $\ell-1$ encountered 
counterclockwise around the face (see Fig.~\maptoquad\ for an example). 
Corners labeled $\tau_i+1$ are finally 
connected by arches to an extra vertex labeled $\tau_i$ at the center
of the face $i$. The set of all arches forms the edges of the 
original quadrangulation and the added vertices correspond to its
sources.

An important property of the above coding of multiply-pointed 
quadrangulations by well-labeled maps is that it keeps
track of a number of distances in the original map. Indeed, as already
mentioned, all the vertices incident to face $i$ in the well-labeled 
map correspond to vertices of the original quadrangulation for which 
the minimum in \defell\ is reached for $j=i$. This ensures
the crucial property:
\eqn\dincid{d_i(v)=\ell(v)-\tau_i\qquad\hbox{for $v$ incident
to face $i$}\ .}

Finally, the above bijection holds for prescribed values of the delays
but one may as well consider all {\it pairs made of a quadrangulation} with $p$
marked vertices {\it and of a set of delays} satisfying \contdelay. 
In that case, in order to keep a finite number of configurations of delays, 
we must impose an additional condition, say for instance 
$\min_{i=1,\ldots,p}\tau_i=0$. The resulting configurations are clearly 
in bijection with well-labeled maps with $p$ faces and arbitrary integer 
labels such that the minimal label is $1$ and the delay $\tau_i$ associated 
with each source $i$ is recovered via 
$\tau_i=\min_{v\ {\rm incident\ to\ face}\ i} \ell(v) -1$.

\subsec{Application to triply-pointed quadrangulations}

Let us now apply the above results to the case of planar 
quadrangulations with three distinct marked vertices, distinguished
as $1$, $2$ and $3$. Note first that a planar quadrangulation is 
automatically bipartite, as required by the above construction. 
We shall distinguish two cases of marked vertices:
\item{(i)} the generic case where the three marked vertices are
not aligned;
\item{(ii)} the particular case where these three vertices are aligned.
\fig{A planar map with three faces (a), its skeleton map (b)
and its backbone map (c).}{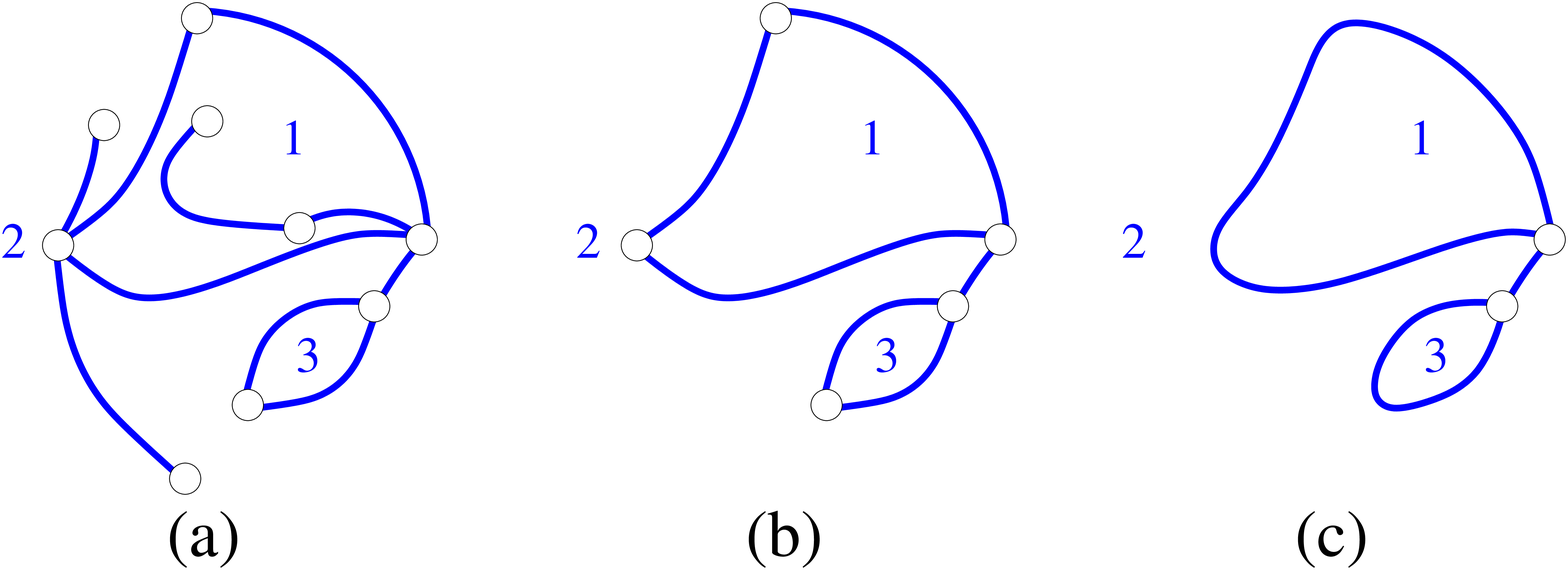}{12.cm}
\figlabel\skeleton
\fig{The seven possible backbones of planar maps with
three faces distinguished as $1$, $2$ and $3$. It will prove
useful to view backbones of type (b), (c) and (d) as degenerate
limits of the backbone (a) when one of its three edges 
is shrunk to a point.}{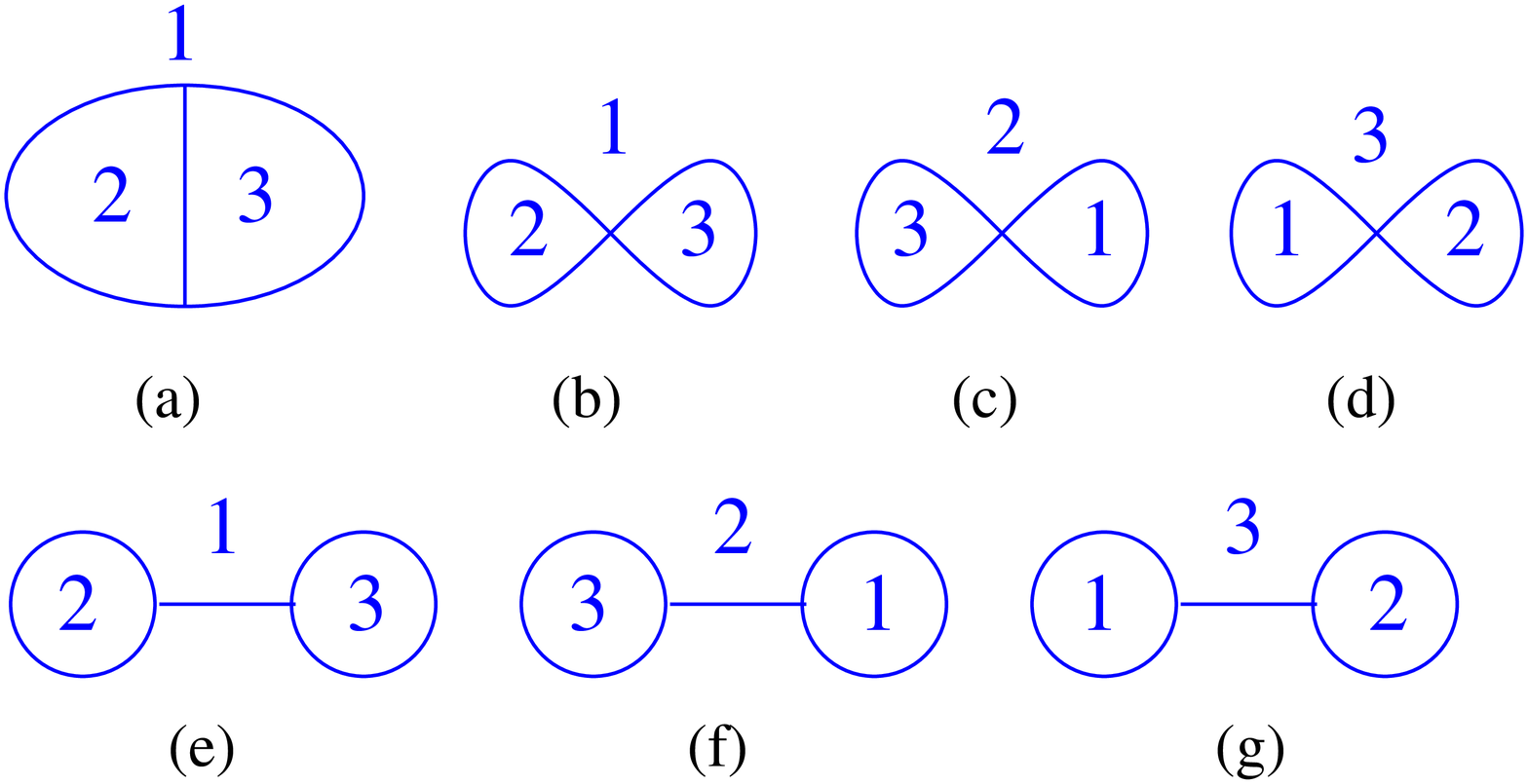}{12.cm}
\figlabel\threemaps
\par \noindent
Let us first concentrate on case (i), leaving the discussion of case
(ii) to the end of this section. Note first that, since the three
points are not aligned, their pairwise distances $d_{ij}$, $1\leq
i\neq j \leq 3$, are all larger than or equal to $2$. Indeed, if at
least one of these distances, say $d_{12}$, was equal to $1$, i.e.\ if
$1$ and $2$ were immediate neighbors, then, from the triangular
inequalities, we would have $\vert d_{31}-d_{23}\vert \leq 1$ and,
from the bipartite nature of the map, this would imply
$d_{31}=d_{23}+1$ or $d_{23}=d_{31}+1$.  In the first case, $2$ would
be on a geodesic path between $1$ and $3$ while in the second case,
$1$ would be on a geodesic path between $2$ and $3$.  In both cases,
the three vertices would be aligned.

Having $d_{ij}\geq 2$ for all $1\leq i\neq j \leq 3$, we may therefore
apply Miermont's construction with $p=3$. The resulting well-labeled map 
is now a planar map with three faces that are distinguished. Such maps can be
classified as follows: it is convenient to view the map as made of 
a {\it skeleton} map decorated by a number of attached tree components 
(see Fig.~\skeleton). 
The skeleton map is obtained by iteratively erasing all 
the edges of the map which are incident to a univalent vertex until no 
univalent vertex is left. The suppressed edges form tree components which, 
once attached to the skeleton, reproduce the original map.
The skeletons themselves can be classified according to their
{\it backbone} map, which is the map obtained by erasing all the bivalent 
vertices of the skeleton and merging their incident edges
into a single edge (see Fig.~\skeleton).
Clearly, by construction, the backbones are planar maps with three
(distinguished) faces
and no univalent nor bivalent vertex. There are only a finite number of 
such maps, all displayed in Fig.~\threemaps. This allows to classify 
all well-labeled maps obtained in the Miermont bijection according to 
their underlying backbone.

The above classification holds for arbitrary planar maps with three
(distinguished) faces and makes no reference to labels or delays. 
For arbitrary fixed values of the delays, the maps in the Miermont
bijection must be equipped with labels satisfying \conlab\ and
one recovers all quadrangulations with three marked vertices at pairwise 
distances compatible with these delays by considering all possible
well-labeled maps with such labels. This requires in
particular to consider all possible backbones in the above classification.

The key point of our derivation that will allow us to keep track of all 
pairwise distances is to impose an extra condition relating the delays
at the sources to the pairwise distances between these sources. 
In other words, we {\it choose  particular values for the delays} 
related to the pairwise distances of three marked vertices on the 
quadrangulation. This particular choice of delays will impose 
extra conditions on the associated well-labeled maps that will restrict 
the possible choice of backbone and imply additional constraints for
the labels. More precisely, let us choose for the delays 
$\tau_1$, $\tau_2$ and $\tau_3$ the values $-s$, $-t$, $-u$ 
with $s,t,u$ as in Eq.~\invpara, namely
\eqn\choicedelay{\eqalign{
\tau_1 &= -s = {-d_{12}+d_{23}-d_{31} \over 2}\ , \cr
\tau_2 &= -t = {-d_{12}-d_{23}+d_{31} \over 2}\ , \cr
\tau_3 &= -u = {d_{12}-d_{23}-d_{31} \over 2}\ . \cr
}}
Note that $s$, $t$ and $u$ are strictly positive integers since the
three points are not aligned, so that the delays are strictly negative.
We have $\vert \tau_1-\tau_2 \vert = \vert d_{23}-d_{31} \vert \leq d_{12}$
from the triangular inequality and again, since we assumed
that the points are not aligned, we cannot have equality. 
This property and the same properties 
under cyclic permutations of the indices ensure that the first condition
in \contdelay\ is satisfied. Moreover, $\tau_1-\tau_2+d_{12}= 2 t$ is even
and so are the two similar quantities obtained by cyclic permutations of the 
indices, 
so that the second condition in \contdelay\ is also satisfied. 
 
Let us now analyze in more details the constraints on the labels of 
the well-labeled maps that result from this particular choice of delays. 
First, we must impose the general condition \conlab, namely that 
all labels of vertices incident to face $1$ (respectively face $2$ and $3$) 
be larger than or equal to $1-s$ (respectively $1-t$ and $1-u$), this
value being reached by at least one of them. In particular, 
for any vertex $v$ incident to face $1$ (respectively $2$ and $3$), 
the quantity $\ell(v)-\tau_1=\ell(v)+s$ (respectively $\ell(v)+t$
and $\ell(v)+u$) measures the distance from this vertex to 
the source $1$ (respectively $2$ and $3$). 
Let us now consider the {\it boundary} between, say faces $1$ and $2$,
defined as the set of vertices and edges {\it incident to both faces} $1$ 
and $2$.  Note that such boundary is part of the skeleton, which is
the union of all boundaries for all pairs of faces. 
For any vertex on the boundary between 
faces $1$ and $2$, the quantity
$(\ell(v)+s)+(\ell(v)+t) = 2\ell(v)+d_{12}$ measures the length of a 
particular path
joining the source $1$ to the source $2$ and passing through $v$. This
length must be larger than or equal to the distance $d_{12}$ between $1$ 
and $2$, which implies that $\ell(v) \geq 0$. 
The same is true for vertices on the boundary between faces $1$ and $3$ or
between $2$ and $3$. We therefore have the crucial property that 
{\it the labels of vertices on the skeleton are non-negative}.

Now consider a geodesic path on the quadrangulation between sources
$1$ and $2$. This path must necessarily intersect the skeleton of
the well-labeled map at a vertex $v_1$ incident to face $1$
and at a vertex $v_2$ incident to face $2$. We have then
$d_{12}=(\ell(v_1)+s)+
d(v_1,v_2)+(\ell(v_2)+t)$, where $d(v_1,v_2)$ is the distance 
between $v_1$ and $v_2$ in the quadrangulation. As $s+t=d_{12}$,
this implies that $\ell(v_1)+\ell(v_2)+d(v_1,v_2)=0$, and, since $v_1$
and $v_2$ lie on the skeleton, all these terms are non-negative. We 
deduce that $d(v_1,v_2)=0$, i.e.\ $v_1$ and
$v_2$ are identical, and moreover $\ell(v_1)=\ell(v_2)=0$. 
The fact that $v_1=v_2$ requires that the boundary between faces $1$ and $2$
be non-empty, which rules out well-labeled maps whose backbone is of type 
(g) in Fig.~\threemaps. Similarly, backbones of type (e) and (f)
are ruled out by considering a geodesic path between 
$2$ and $3$ or between $1$ and $3$ respectively. We are therefore left with
backbones of the type (a), (b), (c) or (d) only.

In the case of a backbone of type (a), we moreover deduce that
there must be a label $0$ on the boundary between $1$ and $2$.
The same property holds of course for the other boundaries too.
We end up with the following constraint: {\it each of the three boundaries} 
between faces $1$ and $2$, $2$ and $3$, and $1$ and $3$ 
{\it must contain a vertex $v$ with label $\ell(v)=0$}.
The same constraint holds in the case of a backbone of type (b), (c) or (d).
In this case, one of the boundaries is made of a single vertex (which 
also belongs to the two other boundaries and  
corresponds to the four-valent vertex of the backbone or of the skeleton).
We deduce that this vertex must have a label $0$. 

\fig{A schematic picture of a well-labeled map whose backbone is of type (a) 
in the classification of Fig.~\threemaps, and the set of constraints
on its labels for our special choice of delays \choicedelay. This constraints
are ``face constraints'' (in squared boxes) for the vertices 
incident to each face, and ``boundary constraints'' (in ellipses) for 
the vertices on the boundary between two faces (thick lines).
The case of maps whose backbone is of type (b), (c) or (d) is obtained
from this picture by simply shrinking one of the three boundary lines into
a single vertex.}{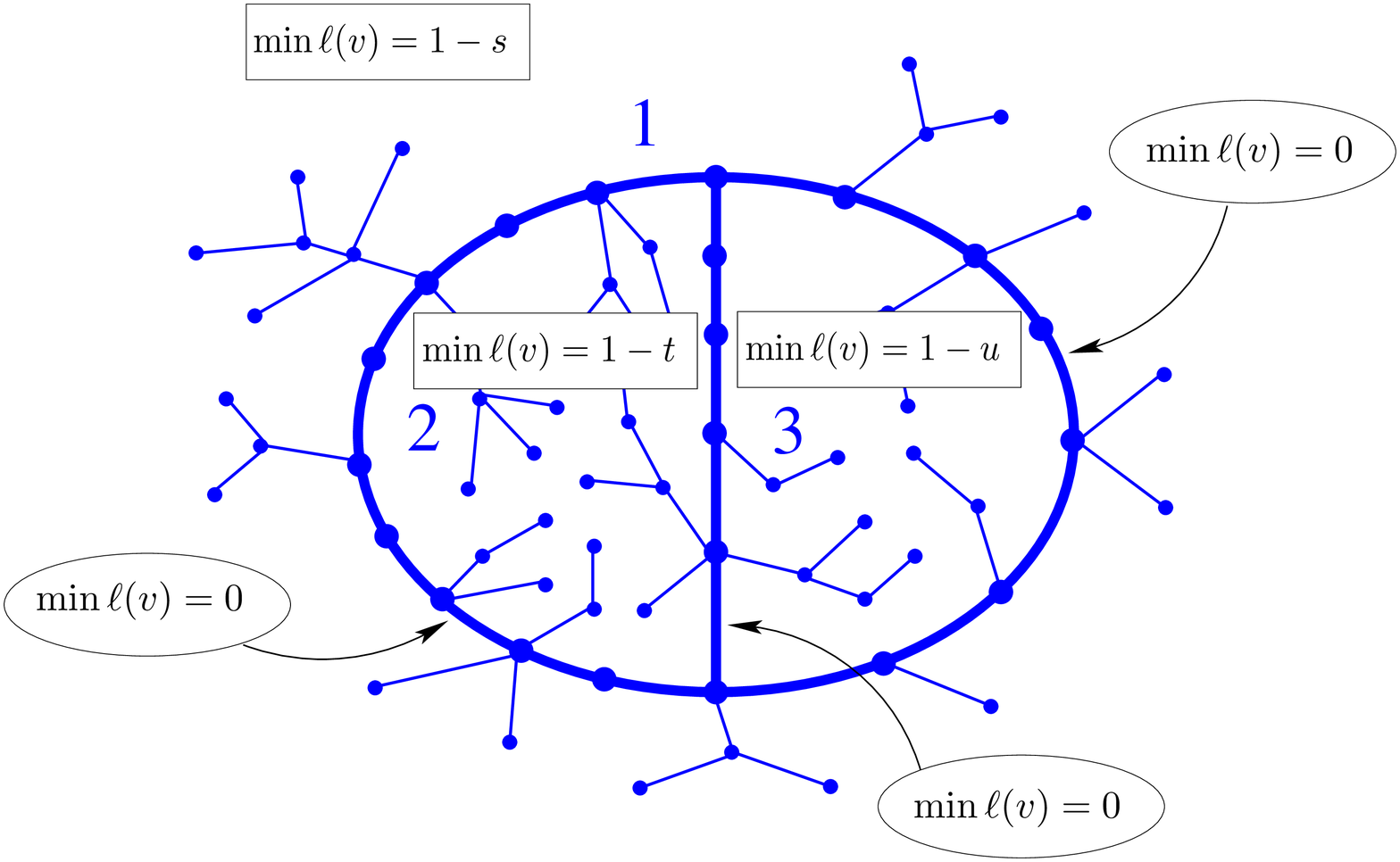}{11.cm}
\figlabel\specialdelays

To summarize the above analysis, the well-labeled maps obtained
for our particular choice \choicedelay\ of delays satisfy
(see Fig.~\specialdelays\ for an illustration):
\eqn\specialcond{\eqalign{&\min_{v\ {\rm incident}\atop {\rm to\ face}\ 1}
\ell(v) = 1-s
\ , \quad \min_{v\ {\rm incident}\atop {\rm to\ face}\ 2}
\ell(v) = 1-t \ ,\quad 
\min_{v\ {\rm incident}\atop {\rm to\ face}\ 3}
\ell(v) = 1-u \ ,\cr
& \min_{v\ {\rm incident}\atop {\rm to\ faces}\ 1\ {\rm and}\ 2}
\ell(v) = 
\min_{v\ {\rm incident}\atop {\rm to\ faces}\ 2\ {\rm and}\ 3}
\ell(v) = 
\min_{v\ {\rm incident}\atop {\rm to\ faces}\ 3\ {\rm and}\ 1}
\ell(v) = 0 \ .\cr}}
By the convention that the minimum of the empty set is $+\infty$, the
last three conditions imply that any two faces have a non-empty
boundary, hence the backbone is necessarily of type (a), (b), (c) or
(d).  Note also that the first three conditions are simply the general
requirement \conlab\ in the Miermont bijection, while the last three
are specific to our choice of delays. For all these conditions to be
compatible, it is crucial that $s$, $t$ and $u$ be strictly positive.

Let us start conversely from a well-labeled map with three
(distinguished) faces whose labels satisfy \specialcond\ for some
strictly positive
integer values of $s$, $t$ and $u$. The labels satisfy in particular
the general condition \conlab\ for delays $\tau_1=-s$, $\tau_2=-t$ and
$\tau_3=-u$, and applying the inverse construction discussed in
section 3.1 will produce a triply-pointed quadrangulation with three
distinct sources (necessarily at a distance larger than or equal to
$2$) and with associated delays $-s$, $-t$ and $-u$. Again a geodesic
path from the source $1$ to the source $2$ will necessarily intersect
the skeleton of the well-labeled map at a vertex $v_1$ incident to
face $1$ and at a vertex $v_2$ incident to face $2$, so that
$d_{12}\geq \ell(v_1)+s +\ell(v_2)+t \geq s+t$. Now, from the fourth
condition in \specialcond, there exists a vertex $v$ at the boundary
between faces $1$ and $2$ with label $\ell(v)=0$ and the union of a
geodesic path from the source $1$ to $v$ and a geodesic path from $v$
to the source $2$ forms a path from $1$ to $2$ of length
$\ell(v)+s+\ell(v)+t=s+t$. We deduce that the distance between the
sources $1$ and $2$ in the quadrangulation is given by $d_{12}=s+t$,
and similarly $d_{23}=t+u$ and $d_{31}=u+s$. In particular, as $s$,
$t$, and $u$ are strictly positive, the sources cannot be aligned.

To conclude, we now have a {\it bijection between, on the one hand, 
triply-pointed quadrangulations whose three marked vertices are not
aligned and have prescribed pairwise distances $d_{12}$, $d_{23}$
and $d_{31}$, and on the other hand, well-labeled maps with 
three faces whose labels 
satisfy \specialcond\ for $s$, $t$ and $u$ related to $d_{12}$,
$d_{23}$ and $d_{31}$ via relation \invpara}.

\fig{A schematic picture of a well-labeled map with two faces (a) 
and a marked vertex $3$ (big circle) as obtained from the
Miermont bijection in the case of aligned marked vertices. 
For our special choice of delays, the constraints on labels
are two ``face constraints'' (in squared boxes) for the vertices 
incident to each of the two faces, one ``boundary constraint'' (ellipse) for 
the vertices on the boundary between the two faces (thick line),
and a ``vertex constraint'' that $\ell=0$ for the vertex 
$3$. Note that this picture can be considered as a degenerate limit
of Fig.~\specialdelays\ when the two boundaries with face $3$ reduce
to a single vertex.}{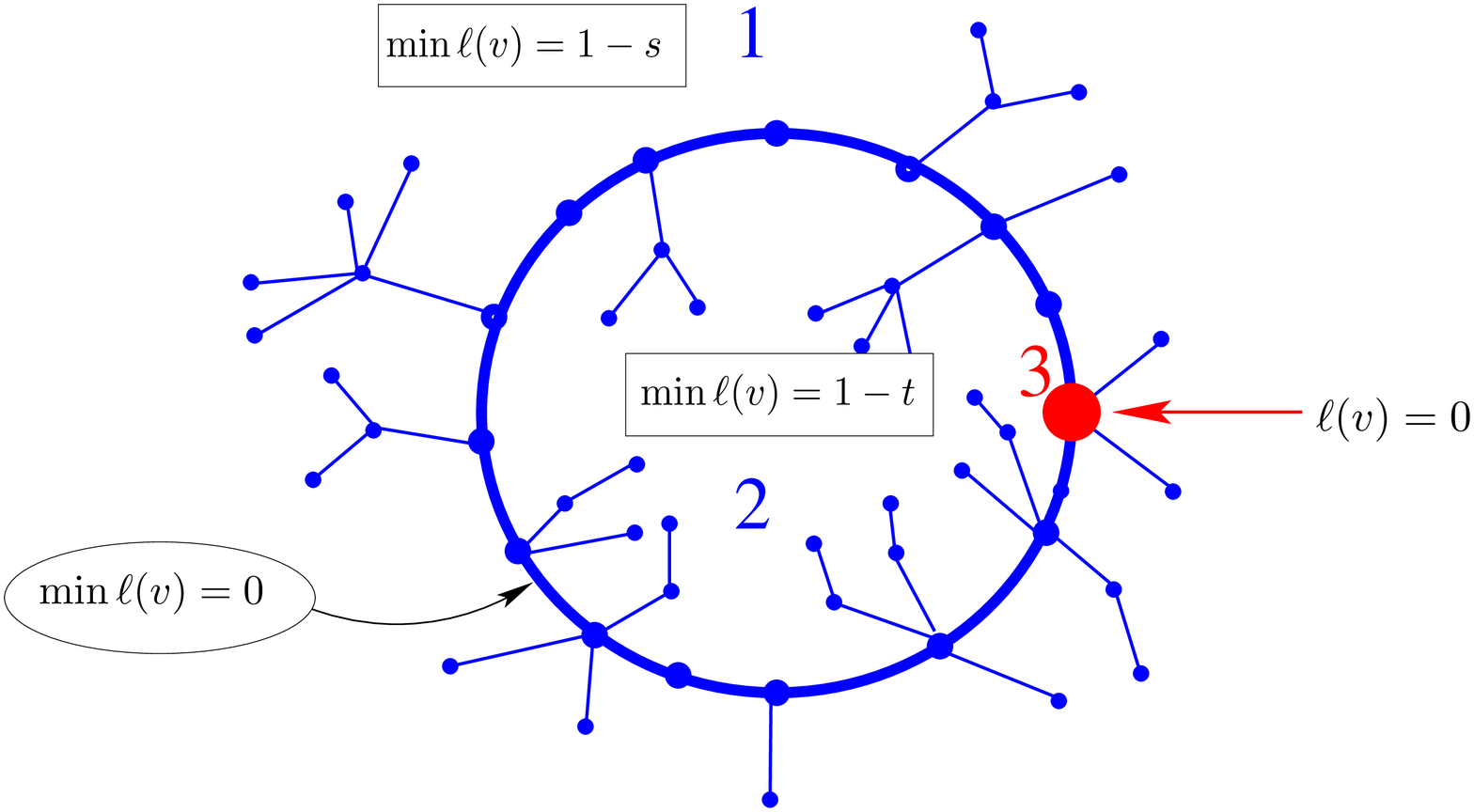}{11.cm}
\figlabel\spectwo

It remains to discuss the case (ii) of aligned marked vertices. 
Let us consider the case where, say $3$ lies on a geodesic
path between $1$ and $2$, i.e. $d_{12}=d_{23}+d_{31}$, 
or equivalently $u=0$. In this case, as the three points are
distinct, we have $d_{12}\geq 2$ and we may apply the Miermont
bijection to our map, now considered as a doubly-pointed quadrangulation 
with sources $1$ and $2$ only, complemented by an extra marked vertex $3$ 
lying on a geodesic between $1$ and $2$. 
For the two sources, we again choose specific delays, namely
$\tau_1=-s=-d_{31}$ and $\tau_2=-t=-d_{23}$. We have 
$\vert d_{31}-d_{23}\vert < d_{23}+d_{31}$ and $d_{31}-d_{23}+ d_{23}
+d_{31}=2 d_{31}$ is even, so that the delays satisfy \contdelay. 
The associated well-labeled map now has two (distinguished) faces and 
its skeleton is a simple
loop, made of all vertices and edges on the boundary of faces $1$ and $2$
(note that the backbone in this case is not strictly speaking
a map as it has no vertex). The same argument as before shows that,
for our particular choice of delays, all labels of vertices on the skeleton
are non-negative and at least one of them vanishes. Now the vertex $3$ 
is necessarily incident to one of the two faces, say face $2$.
Consider then a geodesic path from $1$ to $3$. This path must 
meet the boundary between faces $1$ and $2$ at a vertex $v$ and
we have then $d_{31}=\ell(v)+s+d_3(v)$ where $d_3(v)$ is the distance from
$v$ to the vertex $3$. As $s=d_{31}$, we deduce that $\ell(v)+d_3(v)=0$
and, as both quantity are nonnegative, $\ell(v)=0$ and $d_3(v)=0$.
The vertex $3$ necessarily lies on the skeleton and has the 
minimum label $0$.
We therefore end up with a 
well-labeled map with two (distinguished) faces, with labels satisfying
(see Fig.~\spectwo\ for an illustration)
\eqn\speccondtwo{\eqalign{\min_{v\ {\rm incident}\atop {\rm to\ face}\ 1}
\ell(v) & = 1-s
\ , \quad \min_{v\ {\rm incident}\atop {\rm to\ face}\ 2}
\ell(v) = 1-t  \ ,\cr
& \min_{v\ {\rm incident}\atop {\rm to\ faces}\ 1\ {\rm and}\ 2}
\ell(v) = 0\ , \cr}}
and with a marked vertex $3$ among those vertices of the skeleton 
having label $0$.

Starting conversely from a well-labeled map with two faces, and
labels satisfying \speccondtwo\ for some strictly positive
integer values of $s$ and $t$, and with a marked vertex $3$ on
the skeleton with label $0$, we can use 
the inverse construction to recover a quadrangulation with two sources 
and a marked vertex. All paths between the sources $1$ and $2$ necessarily
cross a vertex $v$ on the boundary between faces $1$ and $2$ so their
length is larger than $(\ell(v)+s)+(\ell(v)+t)$, hence larger than $s+t$.
As the marked vertex $3$ belongs to the boundary and has label $0$, it is
at distance $d_{31}=s$ and $d_{23}=t$ from the sources $1$ and $2$ 
and the distance between the two sources is exactly $d_{12}=s+t$.
In particular, the three points are aligned with $3$ lying between 
$1$ and $2$. 

To conclude, we now have a {\it bijection between, on the one hand, 
triply-pointed quadrangulations whose three marked vertices are 
aligned with, say $3$ lying between $1$ and $2$, and have 
prescribed pairwise distances $d_{12}$, $d_{23}$
and $d_{31}=d_{12}-d_{23}$, and on the other hand, well-labeled maps with 
two faces whose labels 
satisfy \speccondtwo\ for $s=d_{31}$ and $t=d_{23}$, and
with a marked vertex with label $0$ on the skeleton.}

Note finally that the case of aligned vertices is not fundamentally
different from the generic case of non aligned vertices. Indeed,
the well-labeled map with two faces and a marked vertex of
Fig.~\spectwo\ may be viewed as a degenerate form of the generic 
case of Fig.~\specialdelays\ for which 
two of the boundaries, say that between faces $1$ and $3$ and that 
between faces $2$ and $3$ reduce to a single vertex, so that face 
$3$ in practice disappears and its boundary reduces to the single 
vertex $3$. 

\newsec{Enumeration of triply-pointed quadrangulations}

Thanks to the above bijections, enumerating triply-pointed planar
quadrangulations with marked points at prescribed pairwise distances
amounts to enumerating well-labeled maps having three or two faces,
and satisfying \specialcond\ or \speccondtwo\ respectively.  In
practice, the generic case corresponds to a well-labeled map whose
backbone is of type (a) in Fig.~\threemaps. One then recovers all the
other possible cases (backbone of type (b), (c), (d) or aligned
vertices) by simply allowing one or two of the boundaries to reduce to
a single vertex.

In all this section, we shall consider generating functions with a
weight $g$ per edge of the well-labeled map, corresponding to a weight
$g$ per face of the quadrangulation.  Our approach is as follows. In
section 4.1 we recall known results about the enumeration of
well-labeled trees, that is well-labeled planar maps with one face. In
section 4.2 we consider well-labeled maps with two faces, which
correspond to triply-pointed quadrangulations whose marked points are
aligned. Considering this simpler situation first allows us to introduce
some definitions and derive a formula that will be instrumental in 
the next section
4.3, where we consider the general case and derive the expressions
\disctrois\ and \discrformu\ for the three-point function. Finally
section 4.4 is devoted to other applications of our formulas, here in
the ``local limit'' for infinitely large quadrangulations (in
contrast with the continuum limit studied previously).

\subsec{Generating functions for well-labeled trees}

The first building block in our enumeration task is the generating
function $R_i(g)$ of well-labeled trees planted at a corner with label
$i$ and such that all labels are strictly positive. With this
definition we clearly have $R_i=0$ for $i\leq 0$ while, for $i>0$,
$R_i$ satisfies the recursion relation \eqn\recu{R_i={1\over 1-g\,
(R_{i-1}+R_{i}+R_{i+1})}\ ,} where by convention, we set
$R_i\vert_{g^0}=1$ for $i>0$.  This recursion relation was solved in
\GEOD, with the result \eqn\Rdeibis{R_i= R {\q{i}\, \q{i+3} \over
\q{i+1}\, \q{i+2}}\quad\hbox{for}\ i>0 \quad \hbox{with}\ \q{i}\equiv
{1-x^i\over 1-x}\ ,} where the quantities $R$ and $x$ are related to
$g$ by Eq.~\xRtog, and more precisely Eq.~\Rxexplicit.  As already
mentioned in the introduction, the quantities $\log (R_i/R_{i-1})$ for
$i>1$ and $\log(R_1)$ for $i=1$ can be interpreted as the generating
functions for doubly-pointed quadrangulations with two marked vertices
at distance $i$ from each other (and counted with their inverse
symmetry factor). This property is a direct consequence of Schaeffer's
bijection between pointed planar quadrangulations and well-labeled
trees.

Here, from conditions \specialcond\ or \speccondtwo, we will be led to
consider well-labeled trees planted at a corner with some label $\ell
\geq 0$ and such that all labels be larger than, say $1-s$ for some
strictly positive integer value of $s$. The corresponding generating
function is nothing but $R_{\ell+s}$, as obtained by shifting all
labels on the tree by $s$. Note that, in trees counted by
$R_{\ell+s}$, the label $1-s$ may or may not be reached.

\subsec{Generating functions for well-labeled maps with two faces}

\fig{Illustration of the expression for the generating function of
well-labeled maps with two faces, corresponding to triply
pointed-quadrangulations whose marked vertices are aligned. Opening the
skeleton (thick edges) at the marked vertex 3, we obtain a chain with
non-negative labels $(\ell_0,\ell_1,\ldots,\ell_m)$ and with attached
well-labeled trees. Since $\ell_0=\ell_m=0$, the sequence of labels
along the chain forms a Motzkin path. In the enumeration, each
vertex with height $\ell$ in the Motzkin path receives a weight
$R_{\ell+s}$, which counts the possible configurations of the incident
well-labeled tree in face $1$, a weight $R_{\ell+t}$ for the incident
tree in face $2$ and a weight $g$ for the subsequent edge on the
chain.  }{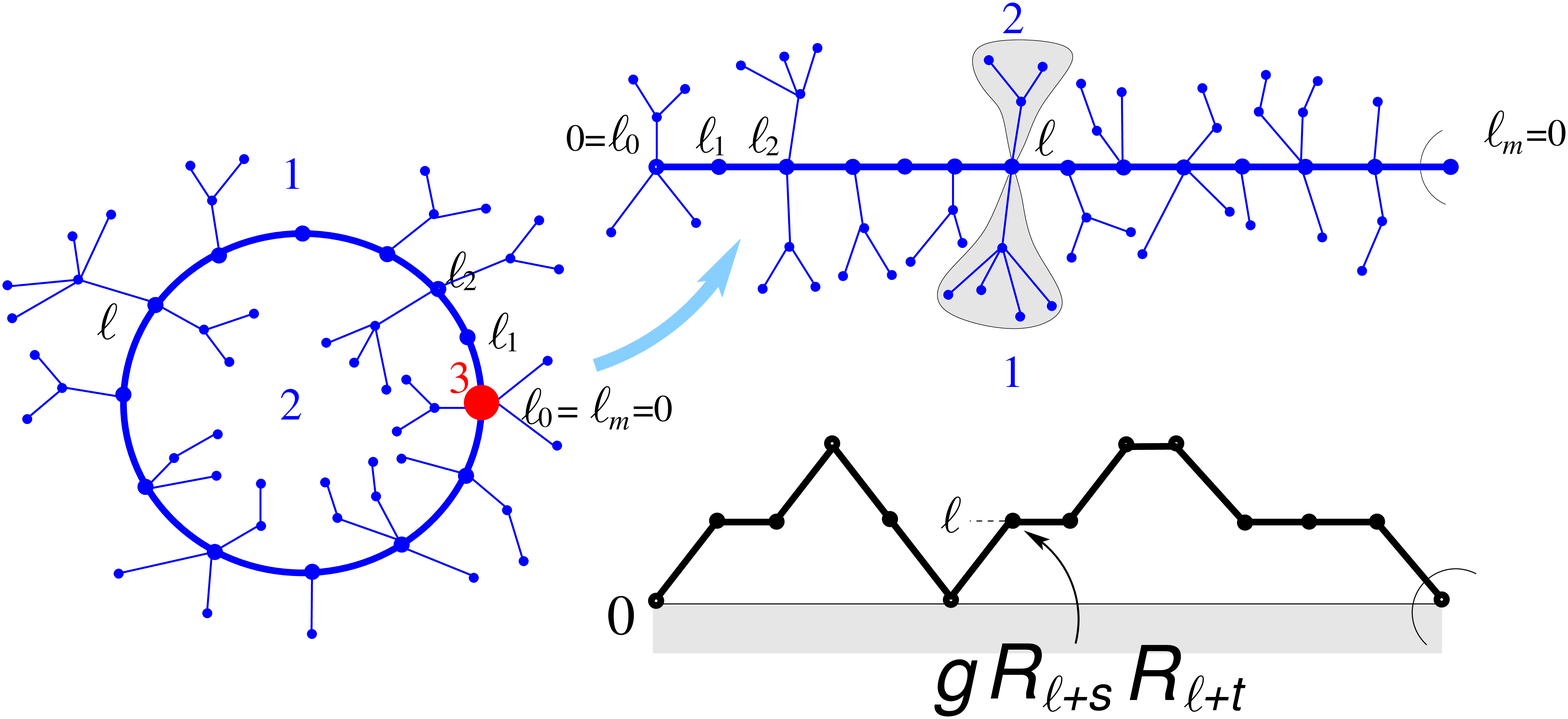}{14.cm} \figlabel\motzkin

The second building block in our enumeration task corresponds to
quadrangulations with three marked aligned vertices. By the bijection
of section 3, we must actually enumerate well-labeled maps with two
faces $1$ and $2$, with labels satisfying \speccondtwo, and with a
marked vertex on the skeleton having label 0.

Consider such a well-labeled map. Its skeleton consists of a single
loop of arbitrary length $m$, and the sequence of labels along it,
read from the marked vertex, can be viewed as the sequence of heights
in a {\it Motzkin path} ${\cal M}$ of length $m$, i.e.\ a path going
from $0$ to $0$ in $m$ steps that never dips below height $0$ and is
made only of $+1$, $0$ or $-1$ steps (see Fig.~\motzkin\ for an
illustration). Let us denote $(0=\ell_0, \ell_1,
\ldots,\ell_{m-1},\ell_{m}=0)$ this sequence. The whole well-labeled
map is uniquely decomposed into the skeleton and its attached tree
components: we see that there are two well-labeled trees attached at
each vertex of the skeleton, one surrounded by face $1$ with all
labels larger than $1-s$, the other surrounded by face $2$ with all
labels larger than $1-t$. The generating function for such trees are
$R_{\ell+s}$ and $R_{\ell+t}$ respectively, where $\ell$ is the label
of the attachment vertex on the skeleton. This leads to consider the
generating function for Motzkin paths with such attached trees:
\eqn\defxst{X_{s,t}\equiv
\sum_{m\geq 0} \quad \sum_{{\rm Motzkin\ paths\ of\ length}\ m \atop
{\cal M}= (0=\ell_0, \ell_1, \ldots,\ell_{m}=0)} \quad
\prod_{k=0}^{m-1} g\, R_{\ell_k+s}\, R_{\ell_k+t}\ ,}
where a weight $g$ is attached to each step of the path (corresponding
to an edge of the skeleton), in addition to the weight $g$ per edge in
the well-labeled trees already counted in $R_i$. Note that there
is no weight attached to the last endpoint ($k=m$) of the path as it
should be identified with the first one ($k=0$), so that we do not
overcount the trees attached there. By convention the path of length
$m=0$ is counted with a weight 1.

Conversely, a Motzkin path with such attached trees can be transformed
back into a well-labeled map with two faces, but this map does not
necessarily satisfy \speccondtwo\ in general, because we have not
imposed that the minimum label in face $1$ (respectively face $2$) be
exactly $1-s$ (respectively $1-t$). Nevertheless, we readily see that
these global constraints are realized by considering the generating
function:
\eqn\ddxst{\Delta_s \Delta_t X_{s,t} = X_{s,t} - X_{s-1,t} - X_{s,t-1}
+ X_{s-1,t-1}}
which corresponds to subtracting the contribution of maps where the
labels in face $1$ are larger than $2-s=1-(s-1)$ or the labels in face
$2$ are larger than $2-t=1-(t-1)$. In the end, using the bijection of
section 3, we find that {\it $\Delta_s \Delta_t X_{s,t}$ is the
generating function for triply-pointed quadrangulations whose marked
vertices are aligned, with pairwise distances $s$, $t$ and $s+t$}.

It turns out that, with the particular form \Rdeibis\ for $R_i$,
$X_{s,t}$ has a very simple expression.  We indeed have the remarkable
combinatorial identity:
\eqn\xst{X_{s,t}={\q{3}\, \q{s+1}\, \q{t+1}\, \q{s+t+3} \over \q{1}\,
\q{s+3}\, \q{t+3}\, \q{s+t+1}}}
which holds for all nonnegative integer values of $s$ and $t$. This
establishes the restricted result \aligned-\fstzero.

Let us discuss how \xst\ can be proved. First, 
from its definition \defxst\ as a sum of weighted Motzkin paths, 
$X_{s,t}$ satisfies the recursion relation 
\eqn\recuxst{X_{s,t}=1+ g R_s R_t\,  X_{s,t}\left(1+ g R_{s+1} R_{t+1}\, 
X_{s+1,t+1}\right)}
obtained by decomposing the Motzkin path according to its first step.
The Motzkin path can be of length $0$ and receives by convention the
weight $1$, or it can start with a $0$ or $+1$ step yielding a first
factor $g R_s R_t$. If it starts with a $0$ step, the rest of the path
is again a Motzkin path with generating function $X_{s,t}$. If is
starts with with a step $+1$, the rest of the path is now a path from
height $1$ to height $0$ which, upon considering the {\it first}
return at height $0$, can be decomposed into a path from $1$ to $1$
with heights larger than or equal to $1$, a step down from $1$ to $0$
and a Motzkin path from $0$ to $0$.  These three components have
respective weights $X_{s+1,t+1}$, $g R_{s+1} R_{t+1}$ and
$X_{s,t}$. Gathering all the possibilities above leads to
Eq.~\recuxst, which uniquely determines $X_{s,t}$ for all $s,t\geq 0$
as power series in $g$, with $X_{s,t}=1+{\cal O}(g)$. Now it is a
straightforward exercise to check that the form \xst\ of $X_{s,t}$
does indeed satisfy \recuxst\ for $R_i$ given by \Rdeibis\ with the
relation \xRtog, and has the expansion $X_{s,t}=1+{\cal O}(g)$. This
completes the proof.

The above argument does not allow to derive the explicit
form \xst\ ab initio from its definition \defxst\ but only checks a 
posteriori that this form matches the definition of $X_{s,t}$. 
A more constructive and historical proof of the formula \xst\ 
is presented in Appendix A where we give two methods for
enumerating so-called ``quadrangulations with a geodesic boundary''
of length $2(s+t)$. Equating the two results precisely 
yields the expression \xst\ for $X_{s,t}$.

Yet another proof is provided by the discussion in section 4.4 below, where
it will be shown that $\Delta_s \Delta_t \log X_{s,t}$ is the generating
function for well-labeled maps with two faces satisfying \speccondtwo,
but without a marked vertex on the skeleton (this induces non-trivial
symmetry factors). By the Miermont bijection, with the particular
choice of delays $\tau_1=-s$ and $\tau_2=-t$, these correspond to
doubly-pointed quadrangulations where the marked vertices are at
distance $s+t$. But the generating function for these doubly-pointed
quadrangulations is given by Eq.\generi\ with $i=s+t$, and we arrive at the
equality:
\eqn\totaltwopre{\log\left({R_{s+t}\over R_{s+t-1}}\right) =
\Delta_s\Delta_t \log X_{s,t} = \log \left( {X_{s,t} X_{s-1,t-1} \over
X_{s,t-1} X_{s-1,t}} \right) \ .}
It is now an easy exercise to derive \xst\ from this equality 
using \Rdeibis\ and the
initial values $X_{s,0}=X_{0,t}=1$.

Finally, let us remark that:
\eqn\Xvsx{\lim_{s,t \to \infty} X_{s,t} = 1+x+x^2 = {x \over g R^2}\,.}
Thus the basic quantity $x$ can be interpreted as a generating
function for Motzkin paths with attached well-labeled trees with no
lower bound on the labels.

\subsec{Generating functions for well-labeled maps with three faces}

We are now ready to enumerate triply-pointed quadrangulations in the
generic case, corresponding to well-labeled maps with three faces $1$,
$2$, $3$ satisfying \specialcond. Let us consider such a well-labeled
map.  As discussed in section 3, its backbone is necessarily of one of
the types (a), (b), (c) or (d) illustrated in Fig.~\threemaps.
\fig{The decomposition of a generic well-labeled map with three
faces into chains and Y-diagrams. We represented the skeleton of the map 
by thick lines and schematized the attached trees by blobs. The skeleton
is the union of the three boundaries $1-2$, $2-3$ and $3-1$ which meet
at vertices $v$ and $v'$. On each boundary, we select the vertices
with label $0$ closest to $v$ and $v'$ (denoted $v_{12}$, $v'_{12}$,
$v_{23}$, $v'_{23}$, $v_{31}$ and $v'_{31}$ with obvious conventions). 
We cut the map at these selected vertices as shown by the dashed lines.
We end up with five pieces in general, namely three chains (in red) and 
two Y-diagrams (in blue). Taking into account the constraints on
the labels of the attached trees (see text), this decomposition 
translates into the relation 
$F(s,t,u;g)=X_{s,t}X_{t,u}X_{u,s}(Y_{s,t,u})^2$.}{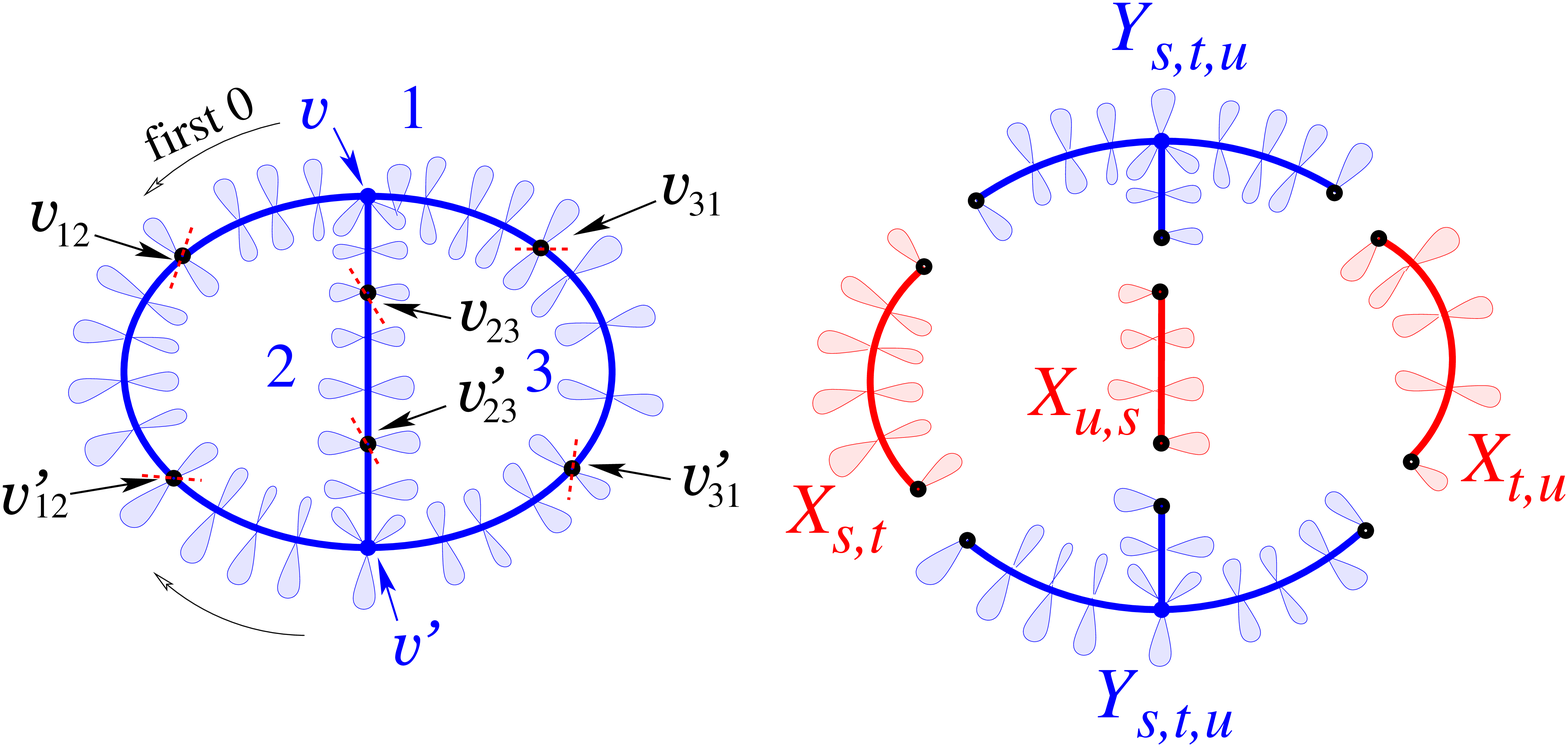}{14.cm}
\figlabel\XY
We first consider the generic type (a), and denote by $v$ and $v'$ the
two trivalent vertices of the backbone, say with $v$ such that faces
$1,2,3$ appear in counterclockwise order around it. Then the skeleton
consists of three chains connecting $v$ to $v'$, corresponding to the
three boundaries $1-2$, $2-3$ and $3-1$. By
\specialcond, the minimal label on each boundary must be 0. Let us
then denote by $v_{12}$ (resp.\ $v'_{12}$) the first vertex with label
0 encountered when following the boundary $1-2$ starting from $v$
(resp.\ $v'$). Similarly we define $v_{23}$, $v'_{23}$, $v_{31}$ and
$v'_{31}$. We can now perform a decomposition of the well-labeled map
by cutting it at the vertices $v_{12}$, $v'_{12}$, $v_{23}$,
$v'_{23}$, $v_{31}$ and $v'_{31}$, as illustrated on Fig.~\XY. More
precisely for each cut vertex, the cutting line is drawn between the
two corners following immediately the two incident skeleton
edges, going counterclockwise around the vertex. 
With this convention, each half of the cut vertex has one
incident skeleton edge, and one attached well-labeled tree. In general
the skeleton is split into five components:
\item{-} three (linear) chains connecting $v_{12}$ to $v'_{12}$, $v_{23}$ to
$v'_{23}$, and $v_{31}$ to $v'_{31}$ respectively,
\item{-} two {\it Y-diagrams}, made of three chains connecting a
central vertex ($v$ or $v'$) to distinct endpoints
($v_{12},v_{23},v_{31}$ or $v'_{12},v'_{23},v'_{31}$).
\par
\noindent There are a number of particular cases. If there is 
a unique vertex with label 0 on one of the boundaries, say $1-2$, then
$v_{12}=v'_{12}$, and the corresponding linear chain can be seen as an
isolated vertex without attached trees, i.e. it is {\it trivial}. If,
say, $v$ has label 0, then $v=v_{12}=v_{23}=v_{31}$, and the
corresponding Y-diagram can also be considered as trivial, made of
an isolated vertex without
attached trees (using our convention to cut within the corners
following immediately the incident skeleton edges counterclockwise, the
three trees attached to $v$ go in different chains). 
For a backbone of type (a), we have $v \neq v'$, hence it is not 
possible that both two Y-diagrams and one of the chains be simultaneously 
trivial. Such situation however occurs when we consider
degenerate backbones types (b), (c), (d), in which case both two Y-diagrams 
and exactly one of the chains are trivial, or when we consider 
well-labeled maps with two faces, for which both two Y-diagrams and 
exactly two of the chains are trivial. It is then convenient to also
include the case of the fully trivial map
reduced to a single vertex, when both two Y-diagrams and three chains
are trivial. 
To summarize, gathering all possible cases, an arbitrary triply-pointed 
quadrangulation
is in correspondence with a well-labeled map that itself can be
decomposed into a combination of three chains and two Y-diagrams
(possibly reduced to isolated vertices).

Let us now investigate the constraints on the labels of attached
well-labeled trees inherited from the first line of \specialcond. 
It is convenient to slightly relax these
conditions and consider well-labeled maps satisfying: 
\eqn\relaxcond{\min_{v\ {\rm incident}\atop {\rm to\ face}\ 1}
\ell(v) \geq 1-s \ , \quad \min_{v\ {\rm incident}\atop {\rm to\ face}\ 2}
\ell(v) \geq 1-t \ ,\quad 
\min_{v\ {\rm incident}\atop {\rm to\ face}\ 3}
\ell(v) \geq 1-u \ .}
We shall call $F(s,t,u;g)$ the generating function of such maps.
Then the generating function for maps satisfying \specialcond\ is
simply given by $\Delta_s\Delta_t\Delta_u F$.
With the relaxed conditions \relaxcond, the constraints on labels
simply factorize into individual constraints for each attached well-labeled
tree.

The case of chains is easy: the sequence of labels along the chain
between $v_{12}$ and $v'_{12}$, say, can be seen as a Motzkin path, 
and all attached trees
previously surrounded by face 1 (resp. 2) have labels larger than
$1-s$ (resp.\ $1-t$). Upon closing the chain, this is the same object as
that considered in the previous section since, for non-trivial
chains, there is only one attached tree at the endpoint $v_{12}$ 
(resp.\ $v'_{12}$), with labels larger than $1-s$ (resp.\ $1-t$). Therefore
the generating function for chains with attached
well-labeled trees satisfying these constraints is nothing but
$X_{s,t}$, including the trivial chain with weight $1$. By the same
reasoning, upon a common cyclic permutation of $1,2,3$ and $s,t,u$, we
find that the generating function for all possible chains between
$v_{23}$ and $v'_{23}$ (resp.\ $v_{31}$ and $v'_{31}$) is $X_{t,u}$
(resp.\ $X_{u,s}$).

We next need to consider the new situation of a Y-diagram, say the one
containing $v$. As said above, it is made of three chains (branches)
connecting the central vertex $v$, to the endpoints
$v_{12},v_{23},v_{31}$. If $v$ has label $0$ then the Y-diagram is
trivial, otherwise every branch has a non-zero length, since
$\ell(v_{12})=\ell(v_{23})=\ell(v_{31})=0$. On each branch, the
sequence of labels can be viewed as a path going from height $\ell(v)$
to 0, made only of $+1$, $0$ or $-1$ steps, and not reaching height 0
before the last endpoint. On the branch to $v_{12}$, every
intermediate vertex has two attached trees, and by \relaxcond\ the one
(within face 1) has labels larger than $1-s$ and the other (within
face 2) has labels larger than $1-t$. The final vertex $v_{12}$ 
has only one attached
tree with labels larger than $1-t$. Similar properties are found for
the other branches, upon a simultaneous cyclic permutation of $1,2,3$ and
$s,t,u$. Finally $v$, being incident to all three faces, has three
attached trees with labels respectively larger than $1-s$, $1-t$ and
$1-u$. Let us denote by $Y_{s,t,u}$ the generating function for
Y-diagrams satisfying such constraints. This
function is determined by the recursive equation:
\eqn\recuystu{Y_{s,t,u} = 1 + g^3 R_s R_{s+1} R_t R_{t+1} R_u R_{u+1}
X_{s+1,t+1} X_{t+1,u+1} X_{u+1,s+1} Y_{s+1,t+1,u+1}}
which is derived in the same spirit as Eq.\recuxst\ for $X_{s,t}$. 
Indeed, let us consider the
label of $v$. If $\ell(v)=0$ then the Y-diagram is reduced to an
isolated vertex with weight 1. Otherwise $\ell(v) \geq 1$, and we cut
the Y-diagram at each first vertex with label 1 encountered when
following each branch starting from $v$ (using the same procedure as
above for dispatching attached trees). We then obtain four pieces :
another Y-diagram (which is trivial iff $\ell(v)=1$) and three
chains. The Y-diagram has labels 1 on its endpoints, otherwise it
satisfies the same constraints regarding the labels on its attached
trees. By decreasing all labels by 1, we find that such Y-diagrams are
enumerated by $Y_{s+1,t+1,u+1}$. The chains can be seen as paths from
1 to 0, that do not reach height 0 before the endpoint. By cutting out
the last step and decreasing all labels by 1, we obtain Motzkin paths
enumerated by $X_{s+1,t+1}$, $X_{t+1,u+1}$, $X_{u+1,s+1}$, for the
respective endpoints $v_{12}$, $v_{23}$ and $v_{31}$. Finally the last
steps contribute with respective weights $g R_{s+1} R_t$, $g R_{t+1} R_u$
and $g R_{u+1} R_s$. Collecting all contributions, Eq.\recuystu\
follows.

We found the remarkably simple form for $Y_{s,t,u}$:
\eqn\ystu{Y_{s,t,u} = {\q{s+3} \q{t+3} \q{u+3} \q{s+t+u+3} \over \q{3}
\q{s+t+3} \q{t+u+3} \q{u+s+3}}}
which can be readily checked by substituting into \recuystu, and
noting that it is the unique solution satisfying $Y_{s,t,u}=1+{\cal
O}(g)$. However we do not have a more ``combinatorial'' derivation for
this formula, similar to those mentioned in the previous section for
$X_{s,t}$. Note that the form \Rdeibis\ for $R_i$ itself still lacks 
such a combinatorial explanation. Alternatively, we may also
write $Y_{s,t,u}$ as a sum:
\eqn\ystualt{\eqalign{& Y_{s,t,u} = \sum_{\ell=0}^{\infty} \tilde{X}_{\ell,s,t}
\tilde{X}_{\ell,t,u} \tilde{X}_{\ell,u,s}\ , \cr
& \tilde{X}_{\ell,s,t}
= {x^{\ell} \q{s+1} \q{s+2} \q{t} \q{t+3} \q{2\ell+s+t+3} \over
\q{s+t+3} \q{\ell+s+1} \q{\ell+s+2} \q{\ell+t} \q{\ell+t+3}}\ , \cr} }
where $\tilde{X}_{\ell,s,t}$ satisfies $\tilde{X}_{\ell,s,t} = g
R_{s+1} R_t X_{s+1,t+1} \tilde{X}_{\ell-1,s+1,t+1}$, and corresponds
to the generating function for branches from $v$ to $v_{12}$ where $v$
has a prescribed label $\ell$.

So far we have derived the generating function for
Y-diagrams containing $v$, however we easily see that Y-diagrams
containing $v'$ have the same generating function $Y_{s,t,u}$ (which
is symmetric in $s,t,u$). Combining the expressions for $X$ and $Y$,
we arrive at the following expression for $F(s,t,u;g)$:
\eqn\lutfin{\eqalign{ & F(s,t,u;g) = X_{s,t} X_{t,u} X_{u,s}
\left(Y_{s,t,u}\right)^2 \cr &\quad = {\q{3} \left( \q{s+1} \q{t+1} \q{u+1}
\q{s+t+u+3} \right)^2 \over \q{1}^3 \q{s+t+1} \q{s+t+3} \q{t+u+1} \q{t+u+3}
\q{u+s+1} \q{u+s+3} } \cr }}
and we readily recognize \discrformu. 
As mentioned above, the desired generating function for
maps satisfying \specialcond\ is $\Delta_s \Delta_t \Delta_u F(s,t,u;g)$. 
By the discussion of section 3, this is precisely the 
three-point function $G(d_{12},d_{23},d_{31};g)$. 
This completes the proof of \disctrois.

\subsec{Local limit for large quadrangulations and statistics of geodesic 
points}

As a final application of our formulas for triply-pointed 
quadrangulations, we can consider the ``local limit'' of
large quadrangulations, obtained by considering the canonical ensemble of 
quadrangulations with $n$ faces and letting $n$ tend to $\infty$, 
{\it keeping the distances between the marked vertices finite}.
As opposed to the scaling limit discussed in section 2.2,
results in this local limit are expected to be non-universal and
specific to quadrangulations,
and would be different for other classes of maps. Again, we can
extract the term $g^n$ of the various generating functions by 
contour integrals in $g$ of the type of Eq.~\canG. In the local
limit of large quadrangulations, these integrals can be evaluated 
exactly by saddle point estimates (see for instance Ref.~\STATGEOD\ 
for a general discussion of this technique). 
By proper normalizations, our enumeration results then translate 
directly into {\it average properties in canonical ensembles} of 
simply- or doubly-pointed large (meaning strictly speaking with 
fixed size $n\to \infty$) quadrangulations, i.e.\ quadrangulations
with one or two marked vertices, here referred to as sources (see again
Ref.~\STATGEOD\ for a discussion on these ensembles).

For instance, from the explicit form \discrformu, we can extract the 
{\it average number} $\langle c(d_{12},d_{23},d_{31})\rangle$ 
of couples of vertices $(2,3)$ at finite distance 
$d_{23}$ from each other and at respective distances $d_{12}$ and $d_{31}$ 
from the source (denoted $1$) in the ensemble of large 
simply-pointed quadrangulations. We find the following formula:
\eqn\smallf{
\eqalign{& \langle c(d_{12},d_{23},d_{31})\rangle =
\Delta_s\Delta_t\Delta_u f(s,t,u)\ ,\ \ {\rm where} \cr
f(s,t,u)&=
{9\over 140}{\left((1+s)(1+t)(1+u)(3+s+t+u)\right)^2
\over (1+s+t)(3+s+t)(1+t+u)(3+t+u)(1+u+s)(3+u+s)}\cr
&\times 
\left(29+20(s+t+u)+5(s^2+t^2+u^2+s t+t u+ u s)\right)\cr
&\times 
\left((s t+t u+ u s +s t u)(4+s+t+u)- s t u \right)\cr}}
and $s,t,u$ are related to $d_{12},d_{23},d_{31}$ via \invpara.
\fig{The ``profile of geodesic points'' $\langle c(s) \rangle_{d}$,
measuring the average number $c$ of vertices lying on a geodesic path
between the two sources, and at distance $s$ from the first source,
in the ensemble of large doubly-pointed quadrangulations with the two 
sources at distance $d$ from each other. This profile is represented 
in (a) versus $s/d$ for $d=10$, $100$ and $1000$ (from bottom to top), 
for all allowed values of $s$. The edge of the profile 
corresponding to values $s=1,\ldots,10$ is represented 
versus $s$ in (b) for $d=10$, $20$ and $50$ (lower values, 
ordered from bottom to top) and in the limit $d\to\infty$ 
(top values in red).}{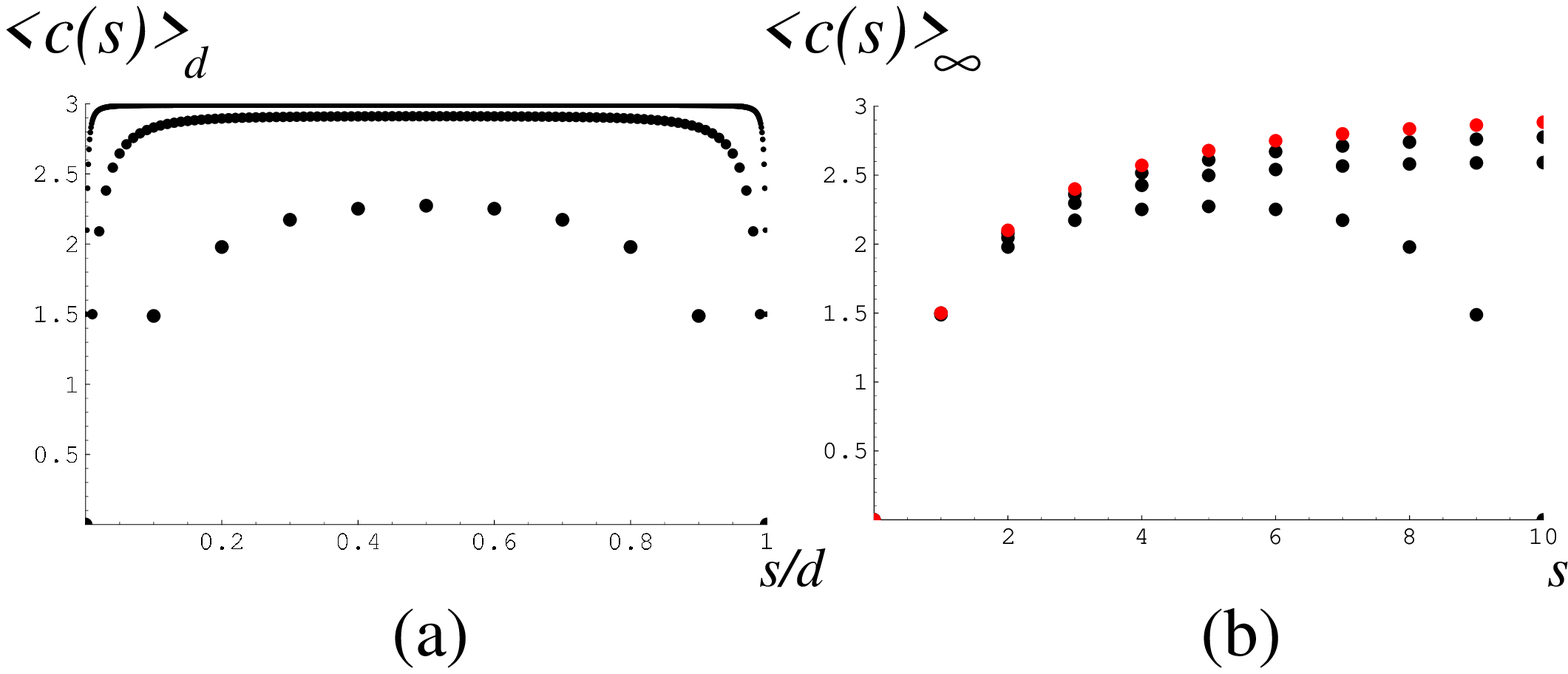}{13.cm}
\figlabel\profilg
\noindent Taking $u=0$, this quantity reduces 
to $\Delta_s\Delta_t f(s,t,0)$ with
\eqn\smallfuzero{\eqalign{f(s,t,0)&=
{9\over 140}{(1+s)(1+t)(3+s+t)
\over (3+s)(3+t)(1+s+t)}\cr
&\times s t
\left(29+20(s+t)+5(s^2+t^2+s t)\right)(4+s+t)\cr}}
and measures, in the ensemble of large simply-pointed quadrangulations,
the average number of pairs made of a first vertex at distance $s+t$ 
from the source and a second vertex
lying on a geodesic path between the source and the first vertex, 
at distance $s$ form the former.
Upon dividing by the known average number $N_{s+t}$ 
of vertices at distance $s+t$ from the source in
large simply-pointed quadrangulations (see for instance Ref.~\MOB\ for the
value of $N_{s+t}$), this gives the average number 
\eqn\densigeo{\eqalign{\langle c(s) \rangle_{s+t}& ={1\over N_{s+t}}
\Delta_s\Delta_t f(s,t,0)\ , \cr {\rm with}\ N_{s+t} &=
{3\over 35}(s+t+1)(5(s+t)^2+10(s+t)+2)\ ,\cr }}
of geodesic points, i.e.\ vertices lying on a geodesic path between 
the two sources, and at distance $s$ from the first one, in the ensemble of 
large doubly-pointed quadrangulations where the two sources
are at distance $s+t$ from each other. Fixing this distance 
to a constant value $s+t=d$ and letting $s$ vary, the 
``profile of geodesic points'' 
$\langle c(s) \rangle_{d}$ versus $s/d$ is represented in 
Fig.~\profilg-(a) for $d=10$, $100$ and $1000$. 
When $t$ becomes large, keeping $s$ finite, we have
\eqn\larget{f(s,t,0)\sim {9\over 28} {s (1+s)\over 3+s} t^4}
and thus
\eqn\largetbis{\Delta_s\Delta_t f(s,t,0)
\sim {9\over 7} {s (5+s)\over (3+s)(2+s)} t^3}
which compared to the average number $(3/7)t^3$ of vertices 
at a large distance $(s+t)\sim t$, gives an average
number 
\eqn\avegeo{\langle c(s)\rangle_\infty={3 s (5+s)\over (3+s)(2+s)}}
of geodesic points at finite distance $s$ from the first source
when the second source is far apart.
This limiting form is represented in Fig.~\profilg-(b), where it is
compared with the profile at finite values $d=10$, $20$ and $50$ of
the distance between the two sources.
In particular, 
far from the first source, i.e.\ when $s$ becomes large, 
Eq.~\avegeo\ gives an average number of $3$ geodesic points far
away from both sources, as apparent in Fig.~\profilg-(a) and (b). 

Beside its average $\langle c(s) \rangle_{s+t}$, we have access to 
the full probability law for the number $c$ of geodesic points. 
Indeed, consider triply-pointed quadrangulations 
where the three sources are aligned, with the third one lying
on a geodesic path between the first two. As discussed in section 3.2,
these are in bijection with well-labeled maps of the type of
Fig.~\spectwo, with two faces and labels satisfying \speccondtwo.
As already discussed, these maps are counted by $\Delta_s\Delta_t X_{s,t}$,
where $X_{s,t}$ accounts for the closed chain of vertices forming
the boundary of the two faces. Now, for a fixed well-labeled map, the third
marked vertex (vertex $3$ in Fig.~\spectwo) can sit on any of the
vertices of the boundary with label $0$ and the number of geodesic points
is given by the number of such vertices. Maps whose boundary has 
exactly $c$ such vertices are simply enumerated by $\Delta_s\Delta_t 
X_{s,t}^{(c)}$, where
\eqn\czeros{X_{s,t}^{(c)}={1\over c}\left({X_{s,t}-1\over X_{s,t}}\right)^c\ .}
Indeed, the quantity $(X_{s,t}-1)/X_{s,t}=g R_sR_t 
(1+g R_{s+1}R_{t+1} X_{s+1,t+1})$ is the generating function for weighted 
Motzkin paths having all intermediate heights strictly above $0$ and
$X_{s,t}^{(c)}$ simply counts cyclic sequences of exactly $c$ such paths.
In particular, we have the consistency relation
\eqn\solc{\sum_{c\geq 1} c X_{s,t}^{(c)}=X_{s,t}}
since in $X_{s,t}$, one of the $c$ vertices with label $0$ is marked on
the chain.
We also have the normalization
\eqn\normc{\sum_{c \geq 1} X_{s,t}^{(c)} = \log X_{s,t}}
which allows to recover the generating function for 
doubly pointed maps with two marked vertices at distance $s+t$ via
\eqn\totaltwo{\Delta_s\Delta_t \log X_{s,t}\!=\!\log \left( {X_{s,t} X_{s-1,t-1}
\over X_{s,t-1} X_{s-1,t}} \right)\!=\!\log\left({\q{s+t+3}\q{s+t}^2
\over \q{s+t-1}\q{s+t+2}^2}\right)\!=\! 
\log\left({R_{s+t}\over R_{s+t-1}}\right)\ .}
As mentioned in section 4.2, this identity can be used to actually
derive formula \xst. 

Combining \czeros\ and \totaltwo, we deduce that, in the canonical
ensemble of doubly-pointed quadrangulations with $n$ faces whose two
sources are at distance $s+t$ from each other, the probability
$p_{s+t;n}(c,s)$ that there be exactly $c$ vertices lying on geodesics
between the two sources at distance $s$ from the first one reads
\eqn\probac{p_{s+t;n}(c,s)= {
\Delta_s\Delta_t X_{s,t}^{(c)}\vert_{g^n}\over \log(R_{s+t}/R_{s+t-1})
\vert_{g^n}}\ .}
Again we can consider the local limit of large quadrangulations by
sending $n\to \infty$ and keeping $s$ and $t$ finite, leading, after
extracting the $g^n$ terms by a saddle point technique, to
a probability
\eqn\probacinf{\eqalign{p_{s+t}(c,s)& = {1\over N_{s+t}}
\Delta_s\Delta_t {f(s,t,0)\over A_{s,t}^2} \left({A_{s,t}-1\over A_{s,t}}
\right)^{c-1}\cr
{\rm with}\ A_{s,t}&=3{(s+1)(t+1)(s+t+3)\over  (s+3)(t+3)(s+t+1)}\cr}}
and with $f(s,t,0)$ and $N_{s+t}$ as in \smallfuzero\ and \densigeo.
In particular, when $t$ becomes large, we find a probability
\eqn\problim{p_\infty(c,s)= 
{s+3\over 2}\left({2s\over 3(s+1)}\right)^c
-{s+2\over 2}\left({2(s-1)\over 3 s}\right)^c}
that there be $c$ geodesic points at distance $s$ from the first source
if the two sources are far apart.
Finally, far away from the first source, i.e.\ when $s$ becomes large,
this probability reduces to 
\eqn\problimlim{p_\infty(c)= 
{1\over 2}\left({2\over 3}\right)^c\ , }
which is the probability law for the number $c$ of geodesic points 
at any fixed distance, far away from the two sources.

\newsec{Discussion}

In this paper, we computed the generating function for 
triply-pointed planar quadrangulations with three marked
vertices at prescribed pairwise distances. We then derived
its universal scaling form and analyzed its behavior
in various limiting regimes.
The main ingredient in our derivation is the Miermont
bijection between triply-pointed planar quadrangulations with
sources and delays and well-labeled planar maps with three faces.
To keep track of all distances between the marked points, 
we had to supplement this bijection with a particular choice of 
delays, resulting in a more restricted set of well-labeled 
maps amenable to a direct enumeration. The combinatorial 
building blocks in this enumeration were the already known
generating function $R_i$ for well-labeled trees and new
generating functions $X_{s,t}$ and $Y_{s,t,u}$ describing configurations of 
chains or Y-diagrams of such trees. These new functions
have very simple forms \xst\ and \ystu, the first one being also used to 
address the statistics of geodesic points in doubly-pointed 
quadrangulations. We believe that these formulas will be useful
to other enumeration problems related to well-labeled
maps, and consequently to quadrangulations, including maps
with higher genus. In this respect, is it tempting to view 
$X_{s,t}$ as a propagator and $Y_{s,t,u}$ as a vertex, whose combination
into Feynman diagrams builds general well-labeled maps with a number
of constraints in their labels.

A natural question is of course that of the $p$-point function for
$p>3$, which would require to compute the generating function for
multiply-pointed quadrangulations with $p$ sources at prescribed
pairwise distances. Even if the Miermont bijection works in this case
with arbitrary delays, it does not seem possible in general to keep
track of all pairwise distances by a proper choice of delays as we did
for $p=3$. Indeed, there are only $p$ free values for the delays but
$p(p-1)/2$ pairwise distances, so that, when $p>3$, we cannot encode
all pairwise distances in the delays. More precisely, we expect that
our construction will work only if the $p$ marked vertices have
pairwise distances which can be realized as distances between the
centers of $p$ pairwise tangent hyper-spheres in $p-1$ dimensions.  In
other words, this requires that we can write $d_{ij}=s_i+s_j$ for some
set of non-negative integers $s_i$, $i=1,\ldots,p$. 
This is clearly not a generic situation but it includes for instance the 
case of regular simplices, for which
all pairwise distances are the same. 
Multiply-pointed quadrangulations whose $p$ sources have prescribed
pairwise distances of the above restricted form can be enumerated 
explicitly along the
same lines as in the case $p=3$. 
For instance, taking $(s_1,s_2,s_3,s_4)=(s,t,u,v)$ in the case $p=4$, 
one finds a generating function
$\Delta_s \Delta_t \Delta_u \Delta_v F(s,t,u,v;g)$ with 
$F(s,t,u,v;g)=X_{s,t} X_{s,u} X_{s,v} X_{t,u} X_{t,v} X_{u,v}
Y_{s,t,u}Y_{s,t,v}Y_{s,u,v} Y_{t,u,v}$.

Finally, another extension concerns classes of maps more
general than quadrangulations. The discrete two-point function is 
known for instance for bipartite maps with prescribed face degrees
\GEOD\ and has a similar structure as that of Eqs. \generi\ and \Rdei,
but now with $\q{i}$ taking a more general explicit form. 
We expect that the Miermont bijection naturally extends to these 
maps, using the mobile rules of Ref.~\MOB.
We further expect that many of our arguments hold in this case.
In particular, the derivation of $X_{s,t}$ in Appendix A should 
extend to this situation and lead to the same formula \xst, now
with the modified $\q{i}$. As for $Y_{s,t,u}$, we had no combinatorial
derivation but we may still speculate that the form \ystu\ remains
unchanged.

\bigskip
\noindent{\bf Acknowledgments:}
We thank G. Miermont for helpful discussions.
The authors acknowledge support from the Geocomp project, ACI Masse de
donn\'ees, from the ENRAGE European network, MRTN-CT-2004-5616 and from
the Programme d'Action Int\'egr\'ee J. Verne ``Physical applications of
random graph theory''.

\appendix{A}{Derivation of formula \xst\ for $X_{s,t}$}
\fig{A schematic picture (a) of a quadrangulation with a geodesic
boundary of length $2i$. The underlying quadrangulation is symbolized
by the grey background. Its boundary is made of two paths of length $i$
joining two marked vertices (thick circles) which are geodesic
paths in the whole quadrangulation. The configuration can be 
transformed into a true quadrangulation without boundary by adding
winding edges as in (b) or (c). Such quadrangulation can be considered
as a simply-pointed quadrangulation (b) whose source is the top
marked vertex to which we assign a delay $0$. Performing
the Miermont bijection creates a well-labeled tree 
whose structure is displayed in (d).
The quadrangulation can be considered alternatively as a doubly-pointed 
quadrangulation (c) whose sources are the two marked vertices to which we 
assign delays $-s$ and $-t$, (with $s+t=i$). Performing the Miermont 
bijection now creates a well-labeled map with two faces whose structure is
displayed in (e).}{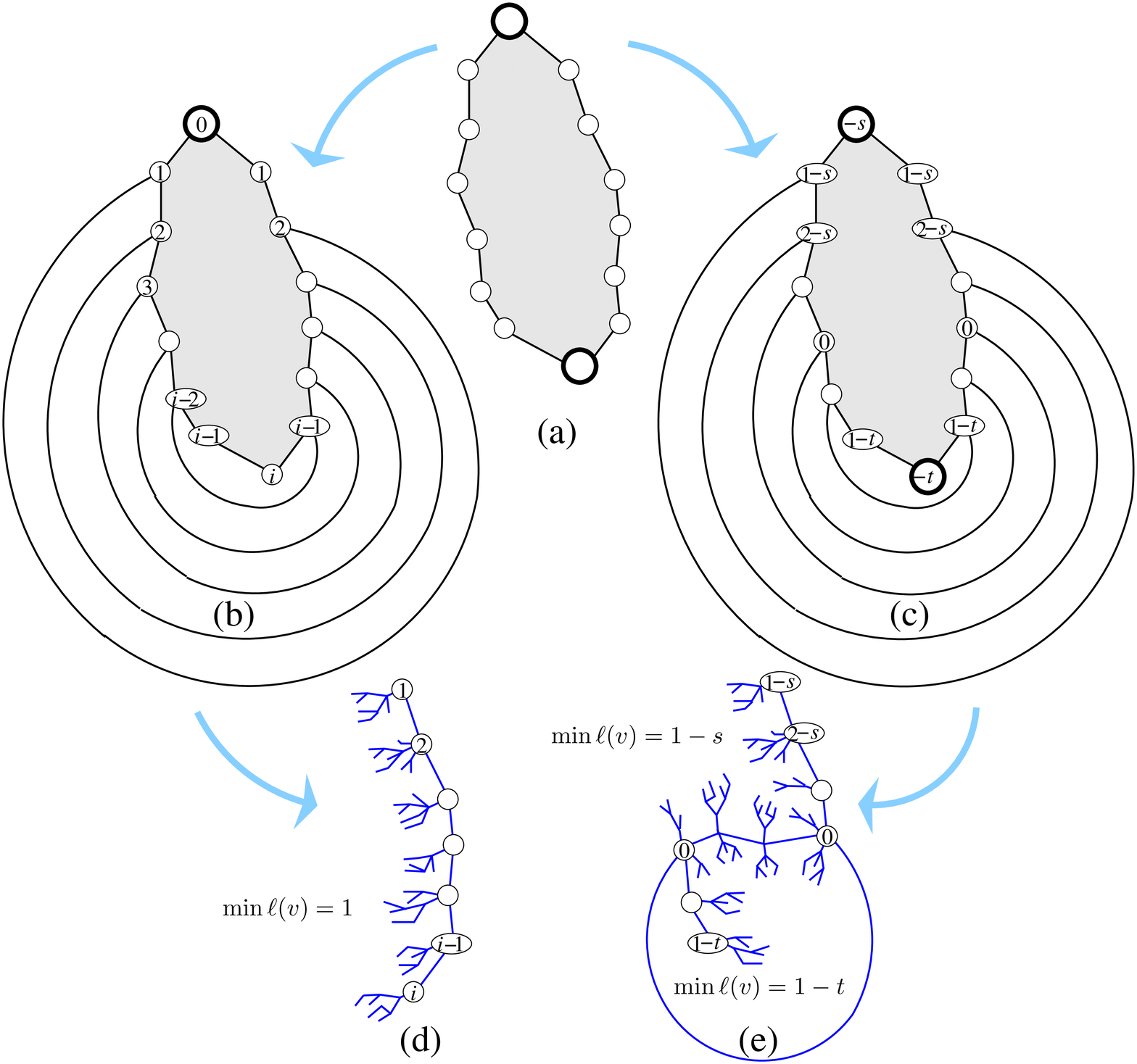}{13.cm}
\figlabel\quadwithb
We give here a derivation of the explicit formula \xst\ for $X_{s,t}$
based on the enumeration of {\it quadrangulations with a geodesic
boundary}. Such objects were introduced and enumerated
in Ref.\STATGEOD\ in the context of a general study of geodesic paths 
in quadrangulations. They can be defined as follows: 
a quadrangulation with a geodesic boundary of length $2i$ ($i\geq 2$) is a
planar map with a marked face of degree $2i$ and with all other faces of 
degree four (squares), and with two marked (and distinguished) vertices 
incident to the marked face {\it at distance $i$ from each other in the map}. 
This last requirement is equivalent to demanding that the boundary 
of the marked face be made of two paths of length $i$ joining the marked 
vertices, which are moreover geodesic paths in the map. 
These paths may possibly meet at common vertices 
(necessarily at the same distance from the marked vertices) 
or even along common edges.
It is convenient, when drawing this map in the plane, to
choose the marked face as the external face. Ignoring this face
results in a quadrangulation having a boundary made of 
the two geodesic paths above joining the marked vertices
(see Fig.~\quadwithb-(a) for an illustration). 

To enumerate these maps, we may transform them into
true planar quadrangulations by adding edges in the external face linking the 
two boundaries of the quadrangulation. We shall refer to these boundaries
as the left and right boundary by viewing the quadrangulation
with the first marked vertex at the top. We then connect by an edge winding
around the quadrangulation each vertex of the right
boundary at distance $k=2,\ldots,i-1$ from the first marked vertex 
to the vertex of the left boundary at distance $k-1$ form the first 
marked vertex (see Fig.~\quadwithb-(b) and (c)). As a result, 
the external face is divided into $i-1$ squares.
Note that the added edges {\it do not modify the distances} to
the two marked vertices of any vertex in the map.
In particular, the two boundaries remain geodesic paths.

We then can use the Miermont bijection to transform this quadrangulation
into a well-labeled map. Let us do this in two ways:
\item{(i)} by considering the map as a pointed map with only one source 
(the first marked vertex) and delay $\tau_1=0$ (this corresponds
to the original Schaeffer construction). This construction
will create a particular type of well-labeled maps with one face,
i.e.\ a well-labeled tree referred to as a ``spine tree'' in
Ref.~\STATGEOD.
\item{(ii)} by considering the map as a doubly-pointed map with two
sources (the two marked vertices) and delays $\tau_1=-s$ and $\tau_2=-t$
for arbitrary strictly positive values of $s$ and $t$ satisfying
$s+t=i$. This will produce a particular type of well-labeled map 
with two faces.
\par \noindent 
In the first case, the label of the vertices of the two boundaries
are their distance to the first marked vertex, ranging from $0$ to 
a maximal value $i$ for the second marked vertex. Applying the construction
of Fig.~\faces, each of the added squares selects an edge of 
the right boundary (see Fig.~\quadwithb-(d)). All the edges of
this boundary but the first one are selected, 
creating a chain of length $i-1$ whose vertices have labels 
$1,2,\ldots,i$. The rest of the quadrangulation gives rise to tree components
that are attached only to the left side of the chain. All these
tree components are well-labeled trees with labels larger
than $\tau_1+1=1$, i.e.\ with strictly positive labels. Note that the 
minimal label $1$ is automatically attained at the first vertex of the chain. 
The generating function for the obtained well-labeled map with the
above structure, counted with a weight $g$ per edge, is simply:
\eqn\firstenu{g^{i-1}\prod_{k=1}^i R_k = g^{i-1} R^i 
{\q{1}\q{i+3}\over \q{3}\q{i+1}}\ .}
Thanks to the Schaeffer bijection, this is also the generating
function for our quadrangulations with a geodesic boundary (up
to a factor $g^{i-1}$ for the spurious added squares).

In the second approach, the labels of the vertices on the 
boundaries are, when going away from the first marked vertex
towards the second marked vertex, $-s,1-s,2-s,\ldots,0$ for the first
$s+1$ vertices and then $-1,-2,\ldots,-t$ for the last $t$.
Applying again the construction of Fig.~\faces, the 
first $s-1$ added squares select edges of the right boundary 
while the last $t-1$ ones
select edges of the left boundary (see Fig.~\quadwithb-(e)). 
The first $s$ edges 
of right boundary but the first one are selected, creating a chain of 
length $s-1$ whose vertices have labels $1-s,2-s,\ldots,0$, while
the last $t$ edges of left boundary but the last one are selected, 
creating a chain of length $t-1$ whose vertices have labels $0,-1,\ldots,1-t$.
Finally, the $s$-th added square is a confluent face that gives rise to
a winding edge connecting the two vertices labeled $0$ on 
the two boundaries. The rest of the quadrangulation creates a
path joining these two vertices with label $0$, and a number
of tree components attached to both side of this path, to the left side 
of the chain of length $s-1$ and to the right side of the chain of
length $t-1$ (see Fig.~\quadwithb-(e)). All these tree components
are well-labeled trees with labels larger than $\tau_1+1=1-s$ or 
$\tau_2+1=1-t$ according to which face they lie in (see Fig.~\quadwithb). 
Note again that the minimal label $1-s$ or $1-t$ is automatically 
attained on the chains. The generating function for the obtained
labeled maps with the above structure is now
\eqn\secondenu{g^{s+t-1}\prod_{\ell=1-s}^0 R_{\ell+s} 
\prod_{\ell=1-t}^0 R_{\ell+t}\times X_{s,t}= g^{s+t-1} R^{s+t}
{\q{1}\q{s+3}\over \q{3}\q{s+1}}
{\q{1}\q{t+3}\over \q{3}\q{t+1}} X_{s,t}\ ,
}
where the factor $X_{s,t}$ comes from the path joining the two vertices
labeled $0$ on the boundaries.
Thanks to the Miermont bijection, this is also the generating 
function for our quadrangulations with a geodesic boundary
of length $2(s+t)$ (up to a factor $g^{s+t-1}$ for the spurious added squares).
Taking $i=s+t$, the formulas \firstenu\ and \secondenu\ thus enumerate
the same objects, hence they must be equal. Equating \firstenu\ 
and \secondenu\ with $i=s+t$ yields 
\eqn\xstbis{X_{s,t}={\q{3}\, \q{s+1}\, \q{t+1}\, \q{s+t+3} \over 
\q{1}\, \q{s+3}\, \q{t+3}\, \q{s+t+1}}}
which is nothing but \xst.

\listrefs
\end